\documentclass[reprint,prd,amsmath,amssymb,nofootinbib,superscriptaddress,floatfix,showkeys,twocolumn]{revtex4-2}

\usepackage{newtxtext,newtxmath}
\usepackage[T1]{fontenc}
\usepackage{graphicx}	% Including figure files
\usepackage{amsmath}	% Advanced maths commands
\usepackage[toc]{appendix}
\usepackage[dvipsnames]{xcolor}
\usepackage{makecell}
\usepackage{float}
\usepackage{xspace}
\usepackage{booktabs}
\usepackage{multirow}
\usepackage{rotating}  % in the preamble
\usepackage{hyperref}
\usepackage{amssymb}
\usepackage{epstopdf}
\usepackage{slashed}
\usepackage{soul}
\usepackage{bm}
\newcommand*{\lya}{Lyman-$\alpha$\,}

\newcommand{\galacc}{GaL$\alpha$CC\xspace}
\newcommand{\lymana}{Lyman-$\alpha$\xspace}
\newcommand{\lymanb}{Lyman-$\beta$\xspace}

\newcommand{\arepo}{\texttt{AREPO}\xspace}
\newcommand{\aida}{AIDA-TNG\xspace}

\newcommand{\aap}{Astron.\ Astrophys.}
\newcommand{\mnras}{Mon.\ Not.\ R.\ Astron.\ Soc.}
\newcommand{\apjl}{Astrophys.\ J.\ Lett.}

\newcommand{\jcap}{JCAP}
\newcommand{\araa}{Ann.\ Rev.\ Astron.\ Astrophys.}

\begin{document}
% \nolinenumbers
% \preprint{APS/123-QED}
%%%%%%%%%%%%%%%%%%% TITLE PAGE %%%%%%%%%%%%%%%%%%%

\title{Gas distributions inside and around haloes in the alternative dark matter simulations AIDA-TNG} 

\author{Chi~Zhang}
\email{chizhang@pmo.ac.cn}
\affiliation{Key Laboratory of Dark Matter and Space Astronomy, Purple Mountain Observatory, Chinese Academy of Sciences, Nanjing 210023, China}
\affiliation{School of Astronomy and Space Science, University of Science and Technology of China, Hefei 230026, China}
\affiliation{SISSA - International School for Advanced Studies, Via Bonomea 265, I-34136 Trieste, Italy}

\author{Enrico~Garaldi}
\affiliation{Kavli IPMU (WPI), UTIAS, The University of Tokyo, Kashiwa, Chiba 277-8583, Japan}
\affiliation{Center for Data-Driven Discovery, Kavli IPMU (WPI), UTIAS, The University of Tokyo, Kashiwa, Chiba 277-8583, Japan.}
\affiliation{IFPU - Institute for Fundamental Physics of the Universe, Via Beirut 2, I-34151 Trieste, Italy} 
\affiliation{SISSA - International School for Advanced Studies, Via Bonomea 265, I-34136 Trieste, Italy} 
\affiliation{INAF - Osservatorio Astronomico di Trieste, Via G. B. Tiepolo 11, I-34143 Trieste, Italy} 

\author{Giulia~Despali}
\affiliation{Dipartimento di Fisica e Astronomia "Augusto Righi", Alma Mater Studiorum Università di Bologna, via Gobetti 93/2, I-40129 Bologna, Italy} 
\affiliation{INAF-Osservatorio di Astrofisica e Scienza dello Spazio di Bologna, Via Piero Gobetti 93/3, I-40129 Bologna, Italy} 
\affiliation{INFN-Sezione di Bologna, Viale Berti Pichat 6/2, I-40127 Bologna, Italy} 

\author{Matteo~Viel}
\affiliation{SISSA - International School for Advanced Studies, Via Bonomea 265, I-34136 Trieste, Italy} 
\affiliation{IFPU - Institute for Fundamental Physics of the Universe, Via Beirut 2, I-34151 Trieste, Italy} 
\affiliation{INAF - Osservatorio Astronomico di Trieste, Via G. B. Tiepolo 11, I-34143 Trieste, Italy} 
\affiliation{INFN - Sezione di Trieste, Via Valerio 2, I-34127 Trieste, Italy} 
\affiliation{ICSC - Centro Nazionale di Ricerca in High Performance Computing, Big Data e Quantum Computing, Via Magnanelli 2, Bologna, Italy} 
 
\author{Lauro~Moscardini}
\affiliation{Dipartimento di Fisica e Astronomia "Augusto Righi", Alma Mater Studiorum Università di Bologna, via Gobetti 93/2, I-40129 Bologna, Italy} 
\affiliation{INAF-Osservatorio di Astrofisica e Scienza dello Spazio di Bologna, Via Piero Gobetti 93/3, I-40129 Bologna, Italy} 
\affiliation{INFN-Sezione di Bologna, Viale Berti Pichat 6/2, I-40127 Bologna, Italy}

\author{Mark~Vogelsberger}
\affiliation{Department of Physics, Kavli Institute for Astrophysics and Space Research, Massachusetts Institute of Technology, Cambridge, MA 02139, USA}

% Abstract of the paper
\begin{abstract}
The nature of Dark Matter (DM) is one of the greatest mysteries of modern astrophysics. While the standard Cold DM (CDM) model successfully explains observations on most astrophysical scales, DM particles have not yet been detected, leaving room for a plethora of different models. In order to identify their observable signatures, we use the AIDA-TNG cosmological simulation suite to predict the distributions of gas and neutral hydrogen (HI) in the CDM, Self-Interacting DM (SIDM), velocity-dependent SIDM (vSIDM), and Warm DM (WDM) models. 
We find that the DM models investigated have very limited impact on the median gas and HI profiles of haloes. 
Nevertheless, for the most massive haloes ($M_{\rm vir}\sim10^{14}\,\mathrm{M}_\odot$), we find that DM self-interactions can reduce the central potential, thereby lowering gravitational heating and the central gas temperature.
We find that, in all models, the halo-to-halo variation in the HI profiles is explained by AGN feedback, and that the specific characteristics of the DM model are largely subdominant. Nevertheless, we detect some systematic difference in the case of SIDM, with more HI surviving close to the centre with respect to other models. We provide fitting functions for the gas and HI profiles. 
We investigate the galaxy-Ly$\alpha$ cross-correlation function (\galacc) for different halo masses, redshifts and observational strategies. We find that at $z=0$ vSIDM can be distinguished from CDM in haloes with $10^{12}\lesssim M_{\rm vir}\lesssim10^{13}\,{\rm M}_\odot$, while SIDM1 can be distinguished from CDM in haloes with $M_{\rm vir}\gtrsim10^{13}\,{\rm M}_\odot$.
We estimate that statistically-robust detection requires sampling $\sim160$ haloes with $\sim20$ sightlines each, a task that can be achieved with current and future facilities like WEAVE, 4MOST, PFS, ELT and WST. 
\end{abstract}

\maketitle
%%%%%%%%%%%%%%%%%%%%%%%%%%%%%%%%%%%%%%%%%%%%%%%%%%

%%%%%%%%%%%%%%%%% BODY OF PAPER %%%%%%%%%%%%%%%%%%

\section{Introduction}
The nature of dark matter (DM) is a fundamental problem in modern physics.
The widely accepted $\Lambda$-Cold Dark Matter ($\Lambda$CDM) model, which assumes that DM is composed of non-relativistic and collisionless particles, successfully explains various large-scale observations by different tracers, including the cosmic microwave background \citep[CMB, e.g.][]{Planck:2018vyg}, the \lya forest~\citep[e.g.][]{SDSS:2004aee,Lidz:2009ca}, weak gravitational lensing and galaxy clustering~\citep[e.g.][]{DES:2017myr,2022PhRvD.105b3520A}.

Despite the great success of the $\Lambda$CDM model, the fundamental nature of DM remains elusive and many candidates remain under intense debate, including e.g. axions \citep{Marsh:2015xka, Graham:2015ouw, 2025arXiv250700705E}, weakly interacting massive particles (WIMPs) \citep{Bertone:2004pz, Feng:2010gw, Roszkowski:2017nbc} and primordial black holes \citep{Carr:2020xqk, Carr:2021bzv}. 
Direct-detection (e.g. XENONnT~\citep{XENON:2018voc}, LZ~\citep{LZ:2022lsv,LZ:2024zvo}, PandaX~\citep{PandaX-II:2017hlx, PandaX:2024qfu}) and indirect-detection (e.g. Fermi-LAT~\citep{Fermi-LAT:2009ihh}, DAMPE~\citep{DAMPE:2017cev}) experiments did not find any significant evidence for the nature of DM.
This, together with the existence of some problems of $\Lambda$CDM at small scales, has powered the exploration of alternative (beyond CDM) models. Among the most studied ones there are the warm dark matter (WDM) and self-interacting dark matter (SIDM) models. 

In the WDM model, it is assumed that DM particles are thermal relics with non-zero initial velocity ~\citep{bode01,viel12}. Their free-streaming length is inversely proportional to the DM particle mass, suppressing fluctuations on scales smaller than such length. 
The tightest constraints on the mass of WDM particles ($>5.7$ keV) currently come from \lya observations in combination with accurate modelling of the non-linear evolution of structure formation in a WDM cosmological model through hydrodynamical simulations~\citep{viel05,Puchwein:2022wvk,Irsic:2023equ, garcia25}. 
Independent constraints come from the number of Milky-Way satellites~\citep{2014MNRAS.442.2487K, 2021JCAP...08..062N, 2022PhRvD.106l3026D}, the properties of stellar streams~\citep{2021JCAP...10..043B, 2021MNRAS.507.1999H} and strong gravitational lensing of both background galaxies and quasars~\citep{2016MNRAS.460..363L, 2020MNRAS.491.4247H, 2020MNRAS.491.6077G}. 

The SIDM models, instead, postulate a new interaction between cold DM particles in addition to gravity, which induces scatterings and an exchange of energy and momentum. This has the net effect of reshaping the matter distributions within haloes through energy transfer from the inner to the outer region~\citep{Spergel:1999mh,Tulin:2017ara}. It is sometimes postulated that the cross-section for such an interaction is velocity dependent (vSIDM model). 
Multiple studies have attempted to constrain the SIDM self-interaction cross-section \citep{Despali:2018zpw,Despali:2022vgq,Eckert:2022qia} under specific conditions. For example, \citet{Eckert:2022qia} find $\sigma/m < 0.19\,\mathrm{cm}^2\mathrm{g}^{-1}$ at $95\%$ confidence level for a collision velocity $v_{\rm DM-DM} \sim 10^3\,\mathrm{km\,s^{-1}}$. These works typically focus on a specific mass scale (e.g. galaxy clusters), enabling more stringent constraints at the cost of generality. 

An alternative approach to direct detection to pinpoint the nature of DM consists in identifying the astrophysical impact of different models and contrasting such predictions with observations. Since the vast majority of DM models are built to comply with available cosmological constraints, galactic and sub-galactic scales are the most promising for this goal. In order to produce reliable predictions on such non-linear scales, cosmological simulations are a fundamental tool. In fact, over the years, several tensions between numerical predictions and observations have been reported at these scales. In addition to the classic `too big to fail'~\citep{2011MNRAS.415L..40B}, `core-cusp'~\citep{1994ApJ...427L...1F,1994Natur.370..629M}, `missing satellite'~\citep{1999ApJ...522...82K,1999ApJ...524L..19M} and the plane of satellite problems~\citep{2018MPLA...3330004P}, strong gravitational lensing recently revealed a new potential inconsistency between CDM predictions and the properties of galactic and cluster satellites \citep{despali2025b,2020Sci...369.1347M}. Thanks to the increased understanding of baryonic physics, growth in simulation resolution and improvement of observations, some of these problems might be solved within the standard $\Lambda$CDM model~\citep[e.g.][]{Simon:2019nxf,2021MNRAS.507.4211E,Font:2020ryo}.

Numerical simulations of beyond-$\Lambda$CDM models are typically DM-only, because of the much leaner computational requirements compared to simulations including gas physics, enabling a wider exploration of the parameter space. Nevertheless, in the past few years a number of hydrodynamical simulations of such models have been published, including \citet{Robertson:2017mgj, Robertson:2018anx, Robertson:2020pxj, Correa:2022dey, Oman:2024kru}
and \citet{2024MNRAS.527.2835S}, which stops at $z=6$.
Recently, a new comprehensive set of simulations named AIDA-TNG \citep{2025A&A...697A.213D} was completed. Compared to previous works, these simulations deliver the first high-resolution, large-volume ($51.7$ and $110.7$ Mpc) cosmological runs with identical initial conditions for the $\Lambda$CDM, WDM, SIDM and vSIDM models. 

In this paper, we employ the new AIDA-TNG simulation suite (briefly introduced in Sect.~\ref{sec:aida_simulations}) to provide a comprehensive study of the total gas and HI content of haloes in the WDM, SIDM and vSIDM models (Sect.~\ref{sec:gas_prop}). We also forward-model observations of the Lyman-$\alpha$ forest and investigate the galaxy-\lya cross-correlation function in Sect.~\ref{sec:CCF}. Finally, we discuss our results and future perspectives in Sect.~\ref{sec:conclusion}.

\section{Simulations}
\label{sec:aida_simulations}
In this section, we briefly introduce the \emph{Alternative Dark Matter in the TNG universe} \citep[AIDA-TNG for short, ][]{2025A&A...697A.213D}, a suite of cosmological magnetohydrodynamical simulations used in this work. All simulations were run with \arepo~\citep{2010MNRAS.401..791S}. They consist of three cosmological volumes of different resolution and size simulated in CDM, three WDM models (particle mass $m_\mathrm{WDM}=1, 3, 5$ keV, referred to as WDM1, WDM3, and WDM5\footnote{Our main analysis only includes WDM3. WDM5 has no hydrodynamical run yet, as its difference from CDM is expected to be small, while WDM1 is used only as a test case at a different resolution.}) and two SIDM models, one (SIDM1) where the interaction cross section is constant $\sigma/m_{\chi} = 1$ cm$^2$ g$^{-1}$, and another (vSIDM) where the cross-section is velocity dependent \citep[following][]{correa21}. In this work, we use the two largest boxes ($L_\mathrm{box} = 110.7$ and $51.7$ Mpc). All simulations start at redshift $z=127$ and adopt the cosmological model from \citet{2016A&A...594A..24P}: $\Omega_{\rm m}=0.3089$, $\Omega_{\rm \Lambda}=0.6911$, $\Omega_{\rm b}=0.0486$, $H_{0}=0.6774$ and $\sigma_{8}=0.8159$. 

For each volume and model, both a DM-only (DMO) and a full-physics (FP) run are available, which will allow us to explore the interplay between alternative dark matter (ADM) models and baryonic physics. The FP runs use the IllustrisTNG galaxy formation model \citep[TNG hereafter]{weinberger17,pillepich18}. The TNG model has been shown to reproduce a broad range of galaxy properties and scaling relations across the galaxy population and as a function of the cosmic epoch. The model updates on the previous Illustris model \citep{vogel14,Torrey14}, including a revised kinetic AGN feedback model for the low accretion state \citep{weinberger17} and an improved parametrisation of galactic winds \citep{pillepich18}. The model includes (i) a sub-resolution treatment of the interstellar medium (ISM) as a two-phase gas where cold clumps are embedded in a smooth, hot phase produced by supernova explosions \citep{2003MNRAS.339..289S}; (ii) feedback from supernova explosions; (iii) the production and evolution of nine elements (H, He, C, N, O, Ne, Mg, Si, and Fe), as well as the tracking of the overall gas metallicity; (iv) density-, redshift-, metallicity-, and temperature-dependent cooling; and (v) super-massive black hole (SMBH) formation, growth and feedback in two different modes (quasar and radio, corresponding to high and low accretion rates, respectively).

Table~\ref{tab:simulations} summarises the properties of the runs used in this work. 
We refer the reader to \citet{2025A&A...697A.213D} and follow-up papers \citep{despali2026,giocoli2026,romanello2026} for a thorough description of the simulations and the models considered.

\begin{table}[h]
  \centering
    \caption{Summary of the AIDA-TNG simulations used in this work and their main properties. From left to right: name of the simulation, linear box size in Mpc, mass of DM and gas particles in solar masses, gravitational softening in kpc.}
    \label{tab:simulations}
    \begin{tabular}{ l c c c c }
    \hline\hline
    \makecell{Name}
      & \makecell{Box\\\lbrack Mpc\rbrack} 
      & \makecell{$m_\mathrm{DM}$\\\lbrack M$_\odot$\rbrack}
      & \makecell{$m_\mathrm{bar}$\\\lbrack M$_\odot$\rbrack}
      & \makecell{$\epsilon_\mathrm{DM,*}^{z=0}$\\\lbrack kpc\rbrack} \\
    \hline
    100/A  & 110.7 & $6.0\times10^{7}$ & $1.1\times10^{7}$ & 1.48  \\
 
    \hline
    50/A   &  51.7 & $3.6\times10^{6}$ & $6.8\times10^{5}$ & 0.57  \\

    \hline\hline
  \end{tabular}
\end{table}

\begin{figure*}
    \centering
    \includegraphics[width=\textwidth]{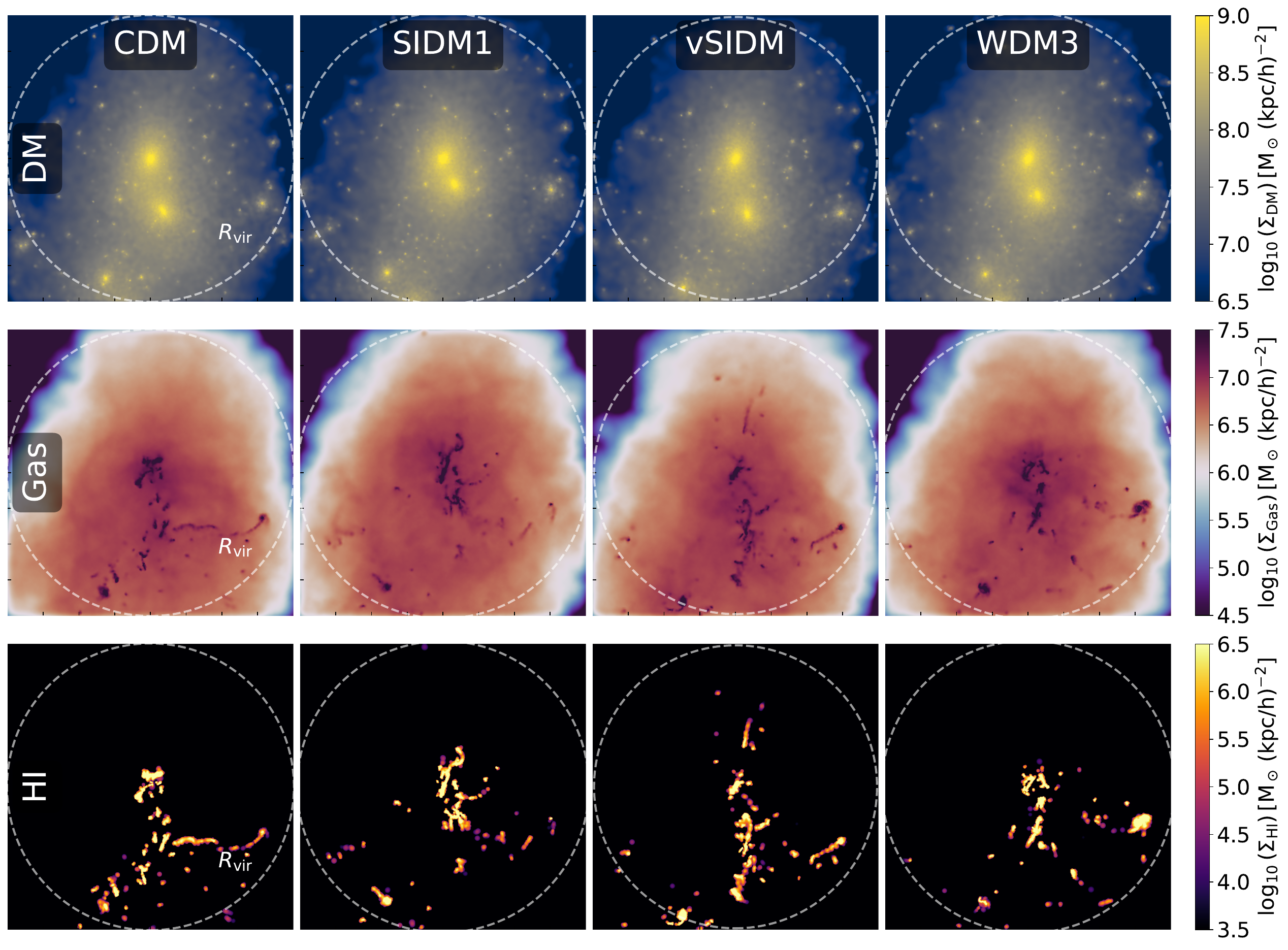}
    \caption{The projected dark matter (top), total gas (middle), and HI (bottom) density for a halo with mass $ M_{\rm vir} \approx 4.3 \times 10^{13}\,\rm M_\odot$ at $z=0$ in the 100/A full-physics runs. We show the same halo in the CDM, SIDM1, vSIDM, and WDM3 (from left to right respectively). The white dashed circle denotes the virial radius of the halo ($R_{\rm vir}\sim400\,\rm kpc/h$). Some small-scale differences can be appreciated between the DM models.
    }
    \label{fig:halo67_proj}
\end{figure*}

\begin{figure*}
    \centering
    \includegraphics[width=\textwidth]{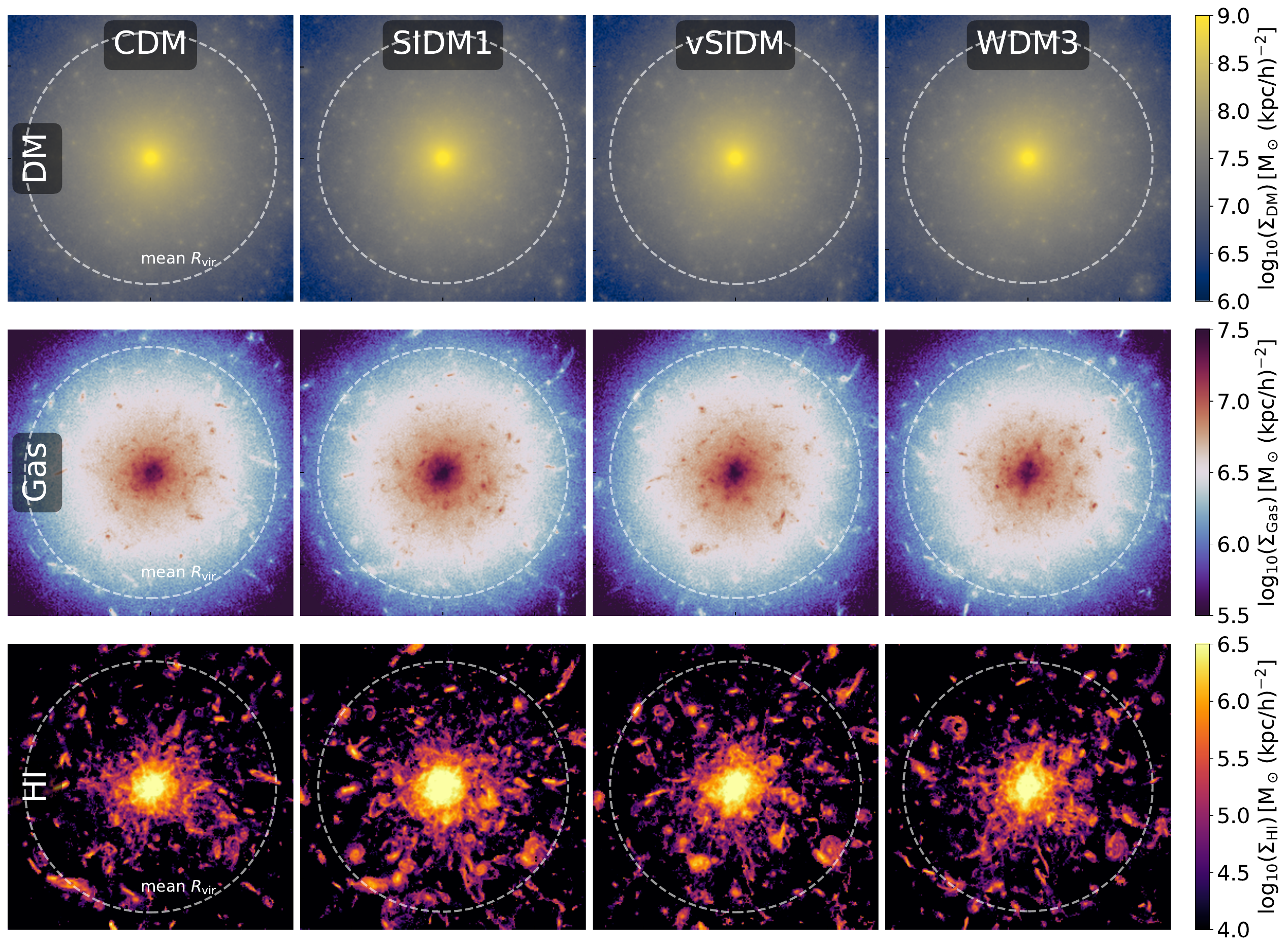}
    \caption{
    Same as Fig.~\ref{fig:halo67_proj}, but stacking of $100$ randomly selected haloes in mass range $\log_{10}M_{\rm vir} \in [12.9, 13.1)\,\rm M_\odot$ at $z=0$, from the 100/A full-physics runs. We use the same haloes for each DM model. The white dashed circle denotes the mean virial radius of the haloes in this bin ($\langle R_{\rm vir}\rangle \sim 273\,\rm kpc/h$). There are no clear differences between DM models.
    }
    \label{fig:stack_100haloes}
\end{figure*}

\section{The gas content of AIDA-TNG haloes}
\label{sec:gas_prop}
We start by investigating the gas content of haloes in different DM models. In order to provide a first visual impression of the gas distribution in the simulations, Fig.~\ref{fig:halo67_proj} shows the projected density of DM (top panels), gas (middle panels) and HI (bottom panels) for a single halo with $M_{\rm vir} \sim 4.3 \times 10^{13}\,\rm M_\odot$ at $z=0$. The halo is extracted from the 100/A full-physics run and is the same halo shown in Fig. 5 of \citet{2025A&A...697A.213D}. Columns from left to right correspond to CDM, SIDM1, vSIDM and WDM3, respectively.
In all DM models, the gas generally follows the underlying DM distribution but remains more diffuse, with localised clumpy structures where cooling is efficient. In turn, the HI distribution mainly follows the gas clumps, with a highly clumpy structure .

We provide a more statistical view of the gas distribution in Fig.~\ref{fig:stack_100haloes}, where we stack $100$ randomly selected haloes (the same for all DM models) within the mass range $M_{\rm vir} \in [10^{12.9},10^{13.1})\,\rm M_\odot$ at $z=0$. As expected, DM and gas form smooth diffuse structures that embed small-scales clumps. Conversely, HI is found only in the compact clumps, approximately following the halo density distribution and with surface densities spanning several orders of magnitude. This is a consequence of complex ionisation (e.g. AGN feedback, star formation, UV background) and self-shielding determining the gas ionisation state \citep{2005ARA&A..43..769V, 2012ARA&A..50..455F, 2012ApJ...746..125H, 2013MNRAS.430.2427R}. We will quantify this clumping in detail in Sect.~\ref{sec:AGN_feedback}.

From this visual inspection, we find that both the central gas and HI distributions in the SIDM1 model exhibit cores that appear larger than in the other models.
We therefore proceed in the next section to a quantitative characterisation of the gas and HI distributions in order to confirm and quantify these visual differences.

\subsection{Halo gas profiles}
\label{sec:gas_profiles}
\begin{figure*}[ht]
    \centering    
    \includegraphics[width=\textwidth]{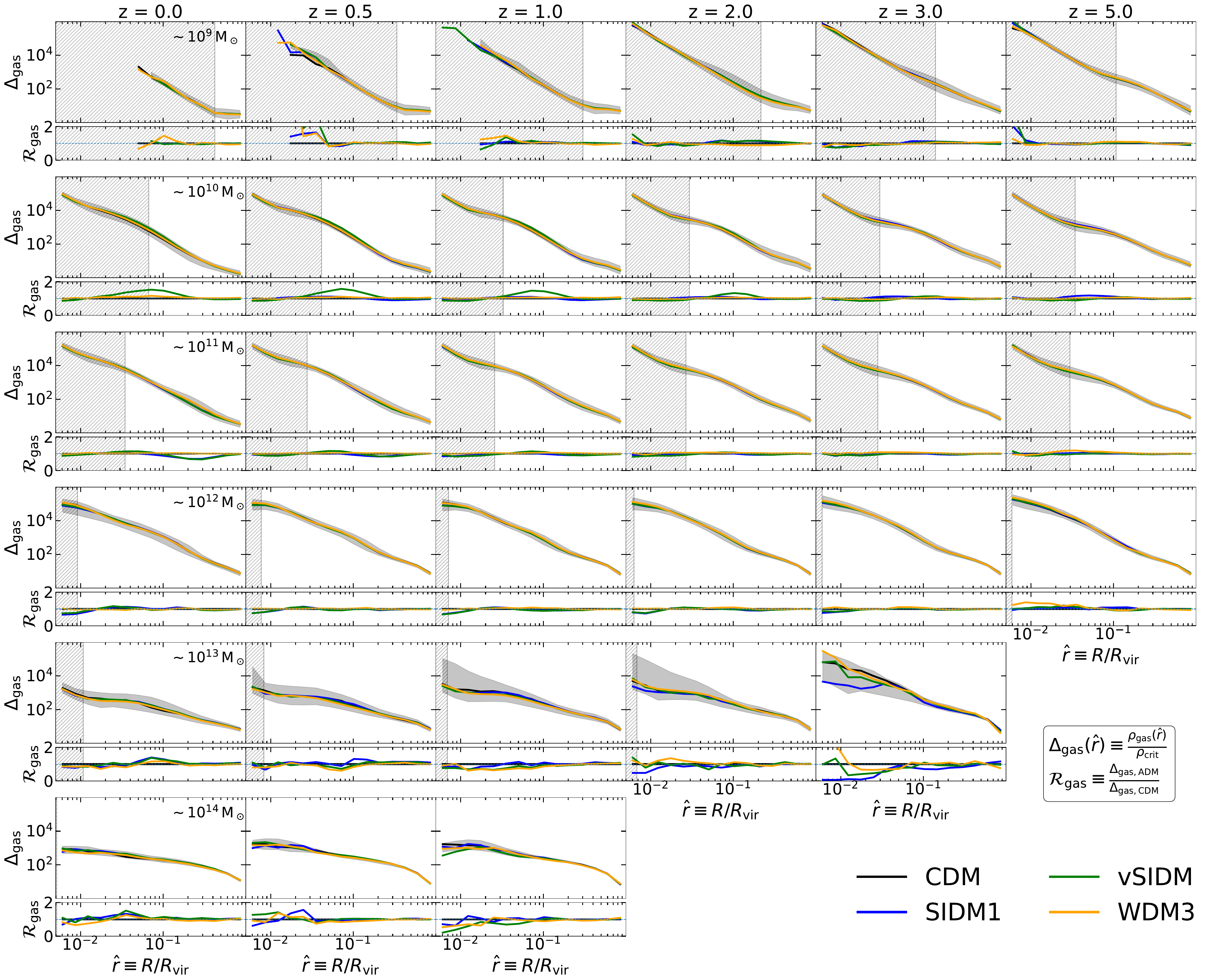}
    \caption{
    The median gas over-density profiles in different DM models, namely: CDM (black), SIDM1 (blue), vSIDM (green), and WDM3 (orange). In the lower part of each panel, we show the the ratio of each profile to the CDM case.
    Columns, from left to right, correspond to increasing redshifts in the range $z=0-5$. Each row corresponds to a different halo-mass bins, increasing from top to bottom. 
    For the top two rows ($10^9,10^{10}\,\rm M_\odot$), we use haloes from the 50/A full-physics runs for better resolution while the rest use the haloes from the 100/A full-physics runs for larger halo samples.
    For each panel, we limit the maximum number of haloes employed to $5000$ for computational efficiency. The exact number of haloes used is reported in Table~\ref{tab:prof_numhalo}.
    The grey band marks the average $68\%$ scatter about the mean profile, while the vertical hatched band shows the estimated resolution limit for the gas distribution in the simulation.
    }
    \label{fig:GAS_profiles}
\end{figure*}
In Fig.~\ref{fig:GAS_profiles} we show the median gas over-density ($\Delta_{\rm gas}$) profiles normalised by the critical density at each redshift (solid lines with different colours for different DM models).

We present over-density profiles for the following reference virial masses:
$M_{\rm vir} \in \{10^{9},\,10^{10},\,10^{11},\,10^{12},\,10^{13},\,10^{14}\}\,\rm M_\odot$.
For each reference mass $\rm M_0$, we restrict the sampled haloes to a narrow bin defined by
$\bigl|\log_{10} M_{\rm vir} - \log_{10} \rm M_0 \bigr| \le 0.1$.
For the top two rows ($10^9,10^{10}\,\rm M_\odot$), we use haloes from the 50/A full-physics runs for better resolution and the rest use the haloes from the 100/A full-physics runs for larger halo samples.
The exact number of haloes used to derive the over-density profiles can be found in Table~\ref{tab:prof_numhalo}.
For each halo, we first normalise the radial distance $R$ to the halo centre by its virial radius as $\hat{r} \equiv R/R_{\rm vir}$, then compute the over-density profile in $25$ logarithmically equispaced bins spanning $\hat{r} \in [5\times10^{-3},1)$.
Subsequently, we compute the median over-density in each radial bin over all sampled haloes. 
We also estimate the $68\%$ central scatter of data for each model, but since they all yield very similar results, we only report the average scatter using a gray-shaded area. 
The panels are organised by halo mass (top to bottom) and redshift (left to right). 
In the bottom part of each panel, we show the ratio of ADM models to CDM. Additionally, since the spatial resolution in the \arepo code is adaptive, we estimate the gas resolution in each halo sample as the median gas cell size within the central $10\%$ of the $R_{\rm vir}$. This is shown by the vertical hatched region in each panel.

First, we examine the redshift evolution of the profiles.
The shape of the profiles does not change significantly with redshift and the underlying DM models, for each halo mass below $10^{13}\,\rm{M}_\odot$.
However, for more massive haloes ($M_{\rm vir} \geq 10^{13}\,\rm{M}_\odot$), a flat central region develops within $\hat{r}\lesssim 0.02$ to $\hat{r}\lesssim 0.04$ as the halo evolves, consistent with the action of a continuous baryonic feedback that reduces the central gas density and enriches the surrounding IGM~\citep{2024A&A...687A.129L}.
In contrast, we do not observe such features at high redshift, suggesting that the impact of feedback process requires significant time to produce a core in the gas distribution. Interestingly, the largest gas over-densities occur in intermediate-mass haloes ($M_{\rm vir} \sim 10^{12}\,\rm M_\odot$), and is reduced in both lower- and higher-mass haloes.
This is likely a consequence of the competition between the accretion of new gas and the ejection of material through feedback mechanisms. Both processes increase in effectiveness with the halo mass, the former thanks to deeper potential wells and the latter as a consequence of more vigorous star formation and black hole activity.

We now focus on the impact of different DM models on the gas over-density profiles. For haloes with $M_{\rm vir} \in [10^9, 10^{12}]\,\rm M_\odot$ (top four rows of the figure), the four DM models do not exhibit significant differences. However, in haloes with $M_{\rm vir} \sim 10^{10}\,\rm M_\odot$, the gas over-density is slightly higher in the vSIDM model than in the others. This is because, in haloes of this mass, the typical relative velocities between DM particles lie in a regime where the velocity-dependent cross section is large. Together with the relatively higher DM density compared to $M_{\rm vir} \sim 10^{9}\,\mathrm{M}_\odot$ haloes, this can modestly strengthen the impact of self-interactions on the gas.

For group-mass haloes ($M_{\rm vir}\sim 10^{13}\,\rm M_\odot$, fifth row in the Figure) we find that, relative to CDM, SIDM1 and vSIDM haloes have slightly lower over-density at $\hat{r} \lesssim 0.06$ at $z=3$, with this offset reducing towards lower redshift. 
Given the small number of simulated haloes of this mass and the large scatter, such feature could be due to sample variance. Larger simulation boxes are needed to conclusively assess this. 

Alternative models, such as WDM and SIDM, alter the DM distribution in the inner regions of haloes. \citet{despali2026} shows that the formation of cores due to self-interactions is reduced in the presence of baryons. However, non-negligible differences remain: a flattening is still present for masses $M_{\rm vir}< 10^{12}\,\rm M_\odot$ in both SIDM1 and vSIDM, while adiabatic contraction driven by the stellar component dominates at the scales of massive galaxies, keeping the profiles cuspy; finally, massive haloes are also cored in SIDM1. It is thus interesting that the changes in the central potential are not strong enough to produce relevant changes in the gas distribution.

Finally, we aim at enabling the use of our results in different modelling approaches \citep[e.g. line intensity mapping, as done in][]{2018ApJ...866..135V}. To this end, we fit the gas distribution with a cored profile of the form:
\begin{equation}
    \Delta_{\rm gas}(\hat{r}) = \frac{\Delta_0}{[1 + (\hat{r}/r_{\rm c})^2]^{3 \beta/2}},
    \label{eq:GAS_fitting}
\end{equation}

where $\Delta_0$ is the normalisation constant, $r_{\rm c}$ is representing the size of the central core, while $\beta$ describes the slope at $\hat{r} \gg r_{\rm c}$. 
We fit each profile in Fig.~\ref{fig:GAS_profiles} (i.e. for each halo mass, redshift and DM model) using a least-square fitting routine\footnote{Specifically, we use the \texttt{scipy.optimize.least\_squares} function in the \texttt{scipy} package \citep{doi:10.1137/0806023}. The fit is performed over the radial bins that have not been excluded, with residuals $\lambda_i=\log_{10}\Delta_{\rm model}(\hat{r}_i)-\log_{10}\Delta_{\rm sim}(\hat{r}_i)$. The minimised quantity is therefore $\sum_i \lambda_i^2$.} and summarise in Table~\ref{tab:gas_fitting} the value obtained for each parameter.

\begin{figure*}
    \centering
    \includegraphics[width=\textwidth]{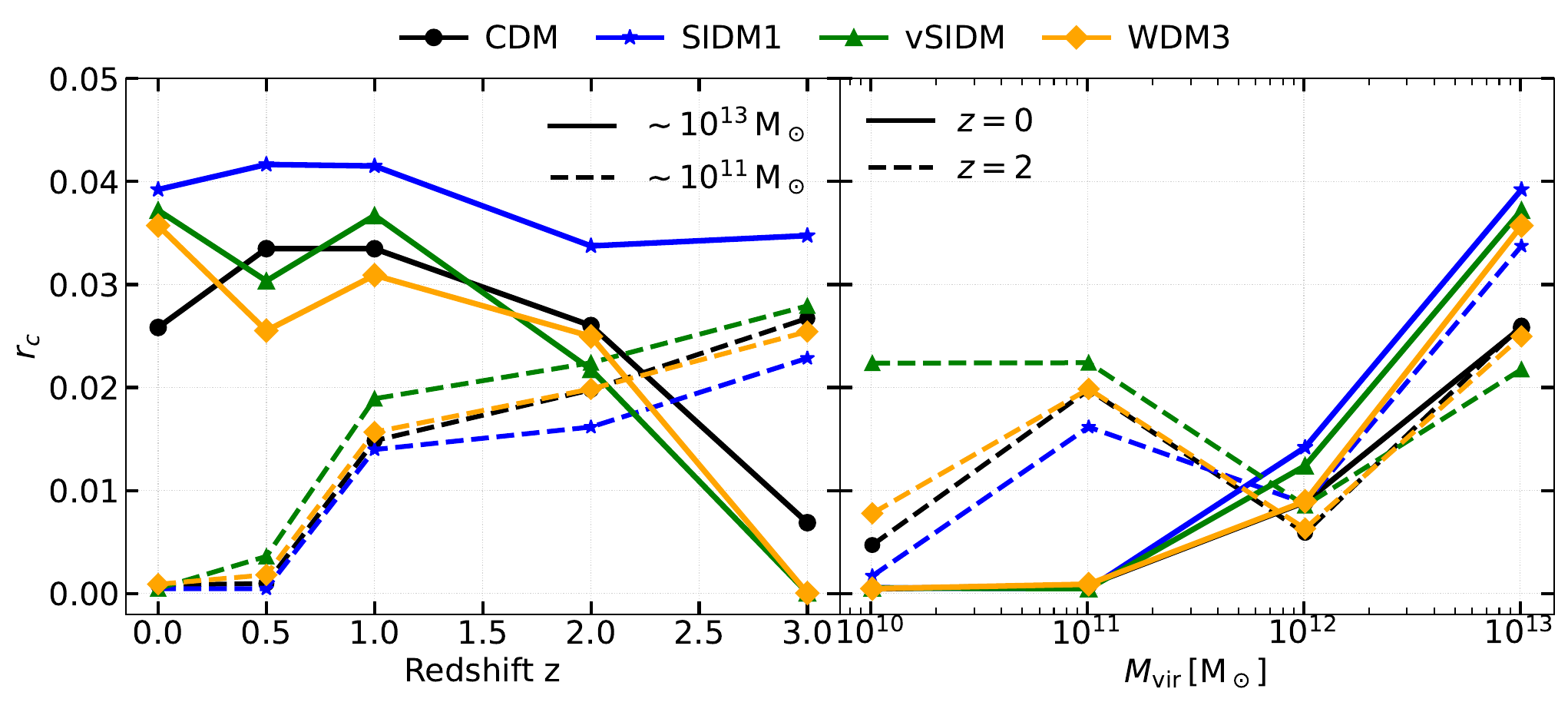}
    \caption{Left panel: Gas core radius $r_\mathrm{c}$ as a function of redshift $z$, for haloes with virial mass $\sim10^{13}\,\rm M_\odot$ (solid lines) and $\sim10^{11}\,\rm M_\odot$ (dashed lines) for different DM models, namely: CDM (black), SIDM1 (blue), vSIDM (green), and WDM3 (orange). Right panel: Gas core radius $r_\mathrm{c}$ as a function of halo mass $M_{\rm vir}$, at $z=0$ (solid lines) and $z=2$ (dashed lines) for different DM models.
    SIDM1 exhibits the largest central gas core, and core sizes in all models increase as redshift decreases. The HI profiles steepen towards lower redshift across all models, though SIDM1 and vSIDM remain slightly flatter.}
    \label{fig:fitting_param_redshift}
\end{figure*}

To move beyond the qualitative comparison of the gas profiles, we also quantify their evolution by examining how the best-fitting parameters of gas profiles depend on redshift, halo mass and DM model.
We present the redshift evolution of $r_{\rm c}$ for haloes with mass $\sim 10^{13}\,\rm M_\odot$ as solid lines in the left panel of Fig.~\ref{fig:fitting_param_redshift} and for haloes with mass $\sim 10^{11}\,\rm M_\odot$ as dashed lines.
For the massive haloes ($\sim 10^{13}\,\rm M_\odot$), we find that the core radius $r_\mathrm{c}$ of the total gas profiles grows with time in all models, although at different rates. 
CDM, vSIDM and WDM3 all show very similar trends of growing gas core. SIDM1, instead, shows a prominent core radius already at $z=3$, accompanied by a much slower growth, so that by $z=0$ its gas core is only slightly larger than for the other models. This suggests that higher redshifts are more promising to differentiate between SIDM1 and other models by looking at the gas core. 
Consistently, SIDM1 introduces the broadest DM core since redshift $z=3$ in massive haloes, due to efficient self-scattering and the DM core grows slowly as the redshift declines.
In contrast, for intermediate-mass haloes ($M_{\rm vir}\sim10^{11}\,\mathrm{M}_\odot$), the gas core radius decreases towards lower redshift, indicating a progressively more centrally concentrated gas distribution. 
In the context of the competitive mechanisms discussed above, this behaviour suggests that, at this halo mass scale, the accretion of new gas increasingly dominates the late-time evolution, resulting a smaller fitted $r_{\rm c}$ towards $z=0$.

In the right panel of Fig.~\ref{fig:fitting_param_redshift}, we show the mass dependence of $r_{\rm c}$ at $z=0$ (solid lines) and $z=2$ (dashed lines). At $z=0$, $r_{\rm c}$ increases with halo mass for all DM models, and the model-to-model differences become more significant for higher-mass haloes. At $z=2$, $r_{\rm c}$ also decreases towards lower halo masses, but exhibits a clear “knee” around $M_{\rm vir}\sim 10^{11}\,\mathrm{M}_\odot$, except in the vSIDM model. Notably, vSIDM maintains relatively large core radii even in low-mass haloes down to $M_{\rm vir}\sim 10^{10}\,\mathrm{M}_\odot$, because their high central densities at $z=2$ place them in a regime of efficient self-scattering that promotes core formation, whereas by $z=0$ the lower central densities move them out of this regime and the effect weakens.

\subsection{HI content of haloes}
\label{sec:hi_profiles}

Gas in haloes can be found in a variety of states. Broadly speaking, hot diffuse gas is challenging to observe, while dense neutral gas can be more easily detected (e.g. through its 21cm lime emission or through absorption of Lyman-$\alpha$ photons). Thus, in order to move one step closer to observable quantities, in this section we turn to the investigation of the neutral hydrogen content of simulated haloes.  

\begin{figure*}[ht]
    \centering
    \includegraphics[width=\textwidth]{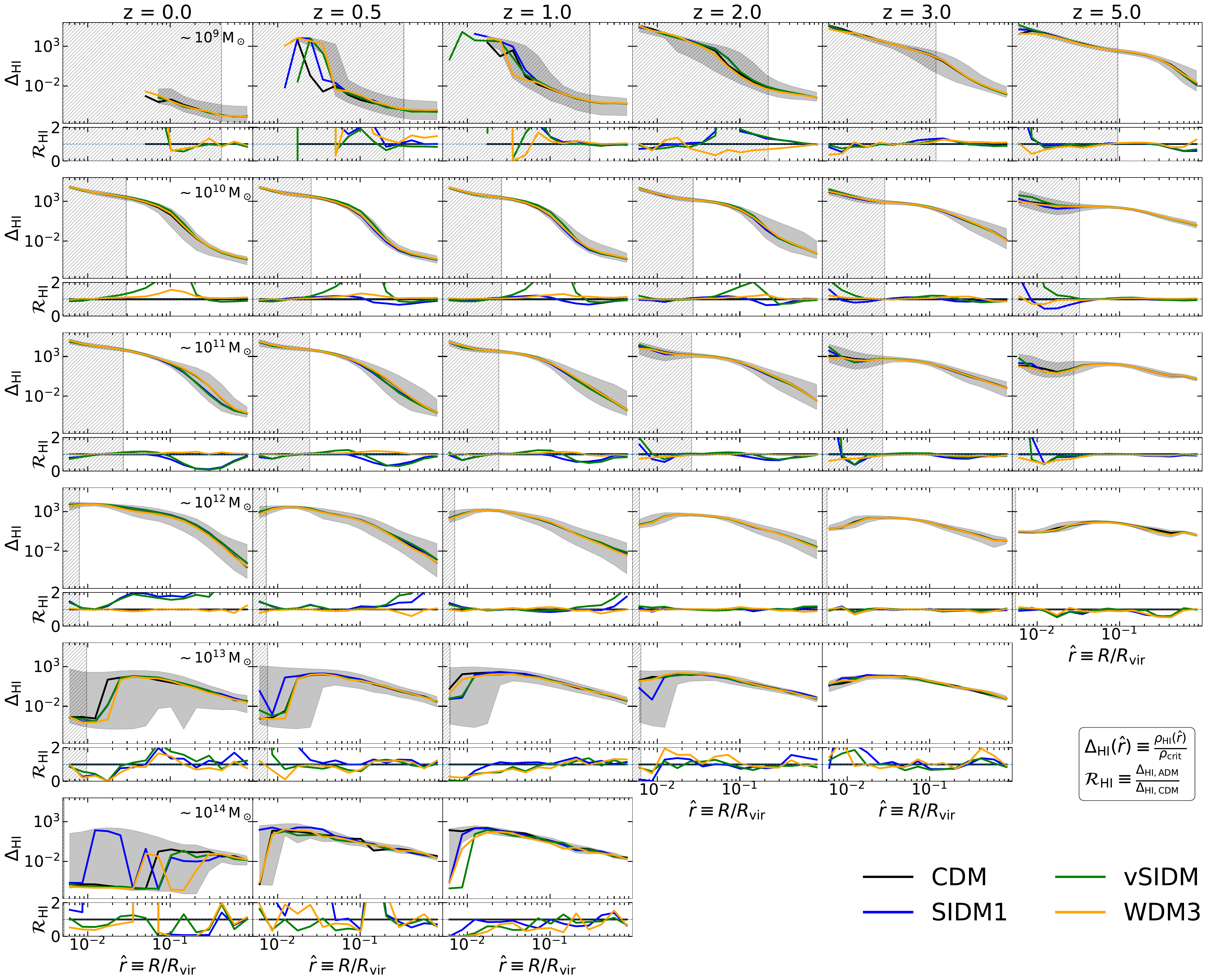}
    \caption{Same as Fig.~\ref{fig:GAS_profiles} but for the HI content of haloes. The HI profiles appear less homogeneous across DM models than the total gas profiles.}
    \label{fig:HI_profiles}
\end{figure*}

To compute the HI profiles, we determine the neutral-hydrogen mass in each gas cell as $m_{\mathrm{HI}} = m_{\mathrm{gas}}\, X_{\mathrm{H}}\, X_{\mathrm{HI}}$, where $X_{\mathrm{H}} \equiv n_{\mathrm{H}}/n_{\mathrm{gas}}$ and $X_{\mathrm{HI}} \equiv n_{\mathrm{HI}}/n_{\mathrm{H}}$. 
Both quantities are available directly from the simulation snapshots. Following the same scheme of Fig.~\ref{fig:GAS_profiles}, we show the profiles and associated scatter in Fig~\ref{fig:HI_profiles}, for the same reference masses and redshifts.

We begin by studying the redshift evolution of the HI over-density profiles. For haloes with $M_{\rm vir} \leq 10^{12}\,\rm M_\odot$ (top four rows), the profiles show a mild radial slope at high redshift and virtually identical for the different DM models. 
As redshift decreases, the central HI over-density increases while the outer over-density declines. This indicates stronger self-shielding in the high density central regions, which preserves cold, neutral gas clumps, whereas gas in the outskirts is more easily ionised by the increasing UV background.
For haloes with $M_{\rm vir} \sim 10^{13}\,\rm M_\odot$ (fifth row), the profiles also show a mild radial slope at high $z$. With cosmic time, however, feedback processes at the centre (mainly AGN feedback, as we will show in Sect.~\ref{sec:AGN_feedback}) increasingly ionise the hydrogen, creating HI-poor central regions. 
The effect is even more evident for the most massive haloes with $M_{\rm vir} \sim 10^{14}\,\rm M_\odot$ (bottom row). 
For both haloes masses, the HI-poor region grows outward as redshift decreases, highlighting the increasing impact of AGN feedback.
The role of AGN feedback is discussed further in Sect.~\ref{sec:AGN_feedback}.

The HI profiles show more variety across dark matter models compared to the gas profiles.
For $10^9\,\mathrm{M}_\odot$ haloes, the differences among DM models in both the central and outer HI over-densities are small compared to the halo-to-halo scatter.
For $10^{10}\,\mathrm{M}_\odot$ haloes, vSIDM shows a significant HI over-density enhancement at intermediate radii ($\hat{r} \simeq 0.03$–$0.2$) relative to the other models.
Haloes with masses $10^{11}\,\rm M_\odot$ and $10^{12}\,\rm M_\odot$ are more strongly sensitive to different DM models. We find clear differences for SIDM1 and vSIDM, which show differences up to a factor of 2 (at $\hat{r} \gtrsim 0.1$ and $z \leq 2$) compared to CDM and WDM3, although of opposite sign for the two mass ranges.
While similar differences can be found also for even larger haloes ($M_{\rm vir} \geq 10^{13}\,\rm M_\odot$), these objects have a much larger halo-to-halo variation that prevent us from faithfully linking them to the underlying DM model.
The increased scatter is driven by the size of this HI-poor region varying from halo-to-halo. 
As we show in the next section, this is due to AGN feedback.

As in the case of the total gas profile, we aim at modelling the gas distribution with simple functions. In order to account for the central HI poor region, we add a hyperbolic tangent function as a multiplicative factor to the profile in Eq.~\ref{eq:GAS_fitting}, i.e.:
\begin{equation}
    \Delta_{\rm HI}(\hat{r}) = \frac{\Delta_0}{[1 + (\hat{r}/r_{\rm c})^2]^{3\beta/2}} \frac{\eta + \tanh[(\hat{r} - r_{\rm tr})/\omega]}{\eta + 1},
    \label{eq:HI_fitting}
\end{equation}
where the extra parameter $r_{\rm tr}$ is the radius where the inner transition happens, $\omega$ is the width of the transition to the halo centre and $\eta$ sets the inner floor and overall strength of the suppression.
We report the fitting parameters in Table~\ref{tab:hi_fitting}.

\subsection{The impact of AGN feedback}
\label{sec:AGN_feedback}
\begin{figure*}
    \centering
    \includegraphics[width=\textwidth]{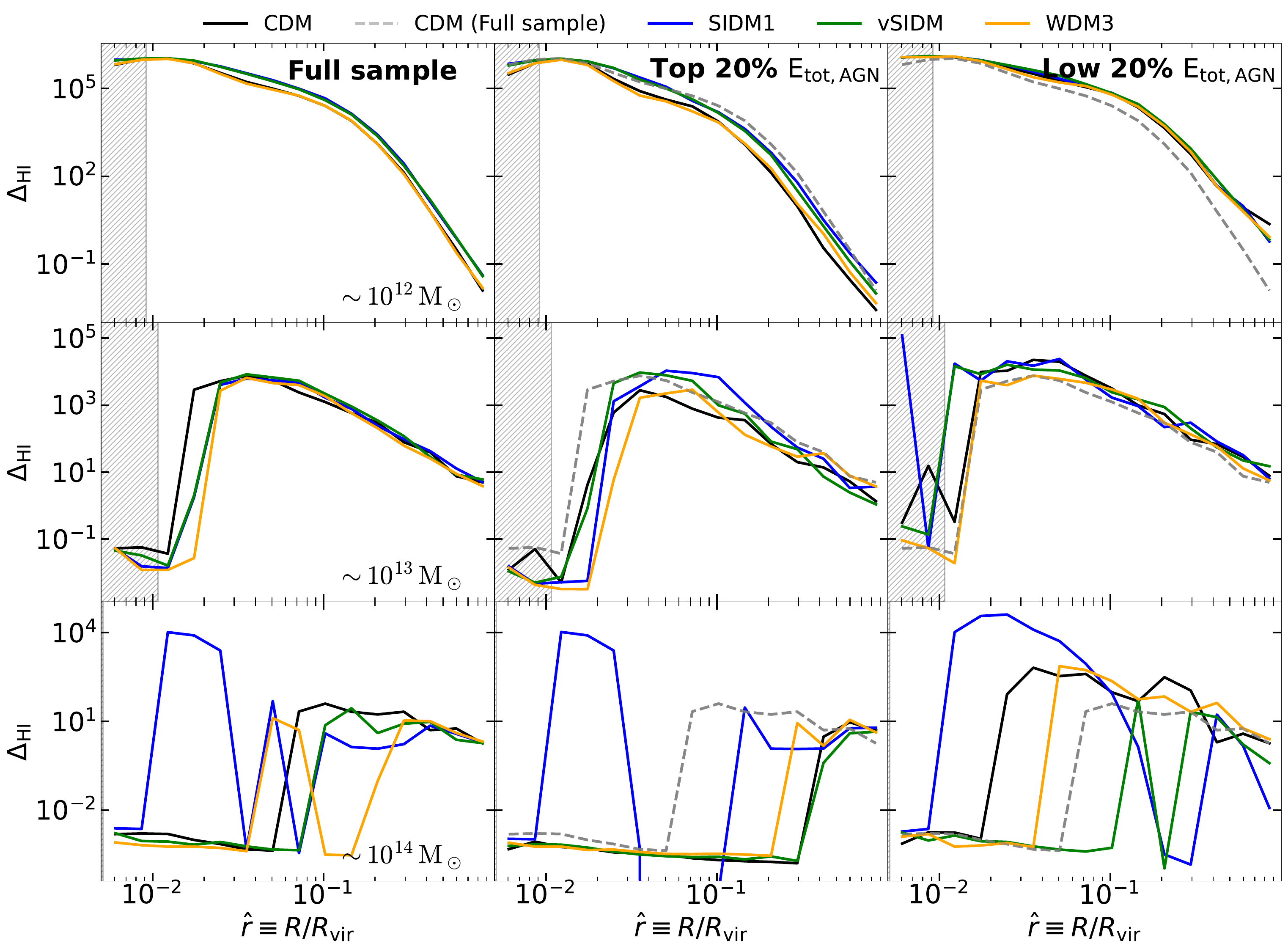}
    \caption{The median HI density profiles at $z=0$ for haloes with masses of $10^{11}$–$10^{12}\,M_\odot$ (top,  $\sim5000$ haloes), $10^{12}$–$10^{13}\,M_\odot$ (middle, $\sim2000$ haloes), and $10^{13}$–$10^{14}\,M_\odot$ (bottom, $\sim210$ haloes). The columns report, from left to right, the profile for the full halo sample (identical to Fig.~\ref{fig:HI_profiles}), for the top $20\%$ and for the bottom $20\%$ of haloes ranked by cumulative AGN feedback injection. Coloured curves correspond to CDM (black), SIDM1 (blue), vSIDM (green), and WDM3 (orange). Grey dashed curves reproduce the CDM full-sample profile of the corresponding mass bin to ease the comparison.
    DM microphysics affects the impact of AGN feedback, resulting in large differences in central HI density.
    }
    \label{fig:HI_profile_BH_egy}
\end{figure*}
The HI distribution in the \aida simulations shows very large `holes' at the centre of haloes with mass $M_{\rm vir} \geq 10^{13}\, \rm M_\odot$ (see Fig.~\ref{fig:HI_profiles} and relative discussion). 
It is now established that AGNs inject large amounts of energy into their surroundings, the so-called AGN feedback, as a byproduct of relativistic accretion of material \citep[e.g.][]{2012ARA&A..50..455F, 2013MNRAS.430.1102T, 2014A&A...562A..21C, 2014ApJ...796....7G}. 
The energy injected can be comparable, or even exceed, the binding energy of gas \citep[e.g.][]{King:2015caa} and therefore can remove part or even all the gas in a halo. Therefore, it is natural to turn our attention to the role of AGN in producing the diversity in HI profiles found in the \aida simulations. 

In the IllustrisTNG galaxy formation model employed in the \aida simulations, AGNs can inject feedback in two different `modes', the so-called quasar and radio modes, active at high and low accretion rates, respectively. In the former, thermal energy is continuously injected in the surrounding gas to simulate radiatively-efficient black hole accretion. In the latter, kinetic energy is injected in discrete events, simulating repeating jet episodes. 
We refer to \citet{2017MNRAS.465.3291W} for details of their implementation. 
For each simulated black hole, the simulation provides the total injected energy in each mode over the life of the black hole ($\mathrm{E}_\mathrm{tot, QM}$ and $\mathrm{E}_\mathrm{tot, RM}$ for the quasar and radio mode, respectively). 
We can then easily define the total injected energy from a black hole as $\mathrm{E}_\mathrm{tot, AGN} = \mathrm{E}_\mathrm{tot, QM} + \mathrm{E}_\mathrm{tot, RM}$. We found that this is an overall better descriptor for the variability of HI profiles, so we only quantitatively discuss this quantity in the following. We note, however, that both $\mathrm{E}_\mathrm{tot, QM}$ and $\mathrm{E}_\mathrm{tot, RM}$ individually correlate with variations at different radii, as qualitatively discussed later. 

In Fig.~\ref{fig:HI_profile_BH_egy}, we show the median HI density profiles at $z=0$ of all haloes (full sample, left column), of the $20\%$ of haloes with the highest (middle column) and lowest (right column) $\mathrm{E}_\mathrm{tot,AGN}$. Profiles are shown for three reference halo masses, namely $10^{12}, 10^{13}, 10^{14}\,\rm M_\odot$, from top to bottom. For comparison, the CDM full-sample profiles are also shown as grey dashed lines in each panel. The shaded regions denote the resolution limit in each case, as defined in Sect.~\ref{sec:gas_profiles}.  For haloes of $10^{12}\,\mathrm{M}_\odot$, the AGN feedback is generally weak, so even the subset with stronger injection of AGN energy does not show a pronounced HI deficit. Still, haloes with higher AGN feedback contain less HI than the overall population, while those with weaker feedback contain more. Also, SIDM1 and vSIDM haloes preserve more HI over the radial range than CDM and WDM3, while in the low-feedback subset the different DM models do not significantly deviate from each other.

We now turn to haloes of $10^{13}\,\rm M_\odot$. In such massive haloes, AGN injects large amounts of energy into the surrounding gas, resulting in a strong suppression of central HI density across all DM models, providing clear evidence that AGN feedback efficiently ionises neutral gas at the halo centre. In the high-feedback subset, we observe even stronger suppression than the CDM full sample reference for all DM models while the opposite happens in the low-feedback subset. For haloes of this mass, the gas and HI distributions are therefore largely shaped by AGN activity, while the influence of the DM potential is secondary.   

\begin{figure*}
    \centering
    \includegraphics[width=\textwidth]{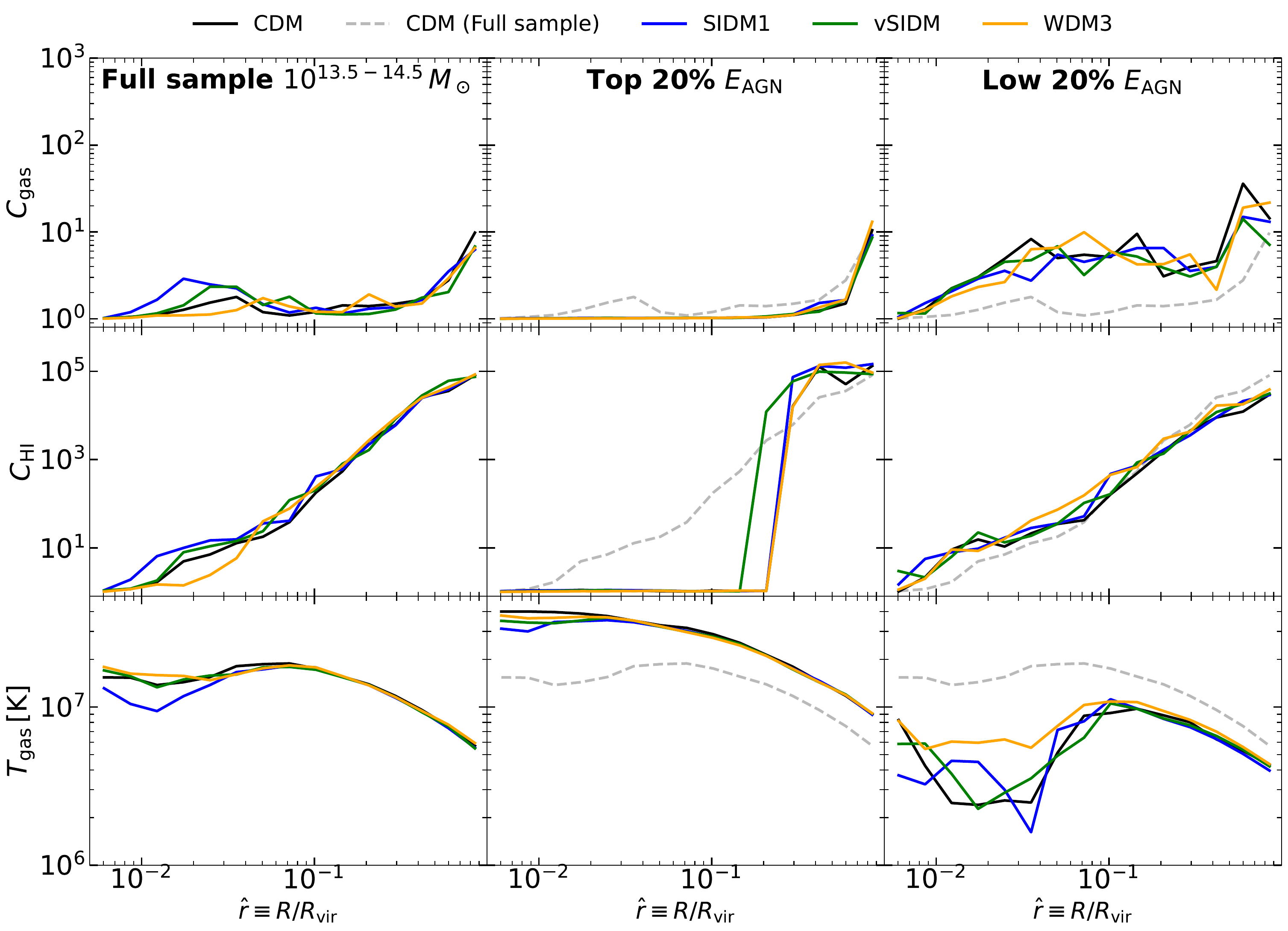}
    \caption{The clumping profiles of gas (top row), HI (middle row) and gas temperature (bottom row) for haloes with mass $\sim 10^{14}\,\rm M_\odot$. The columns report, from left to right, the profiles for the full halo sample (identical to Fig. 4), for the top $20\%$ and for the bottom $20\%$ of haloes ranked by cumulative AGN feedback injection. For SIDM haloes with mass $\sim 10^{14}\,\rm M_\odot$, a relatively low temperature region occurs and increases the clumpiness for total gas and HI.}
    \label{fig:clumping}
\end{figure*}

In the most massive haloes with $M_{\rm vir}\sim10^{14}\,\rm M_\odot$, the number of haloes is low and, therefore, sample variance plays an important role. Nevertheless, the HI distribution seems to be affected by dark matter physics. This is particularly evident in SIDM1, which preserves more HI than other DM models in the central part of the halo. This behaviour is consistent across halo subsamples, making it less likely to stem from sample variance. 
In this case, we find large differences between SIDM1 and vSIDM, originated from the fact that in massive haloes the high DM velocity reduces the velocity-dependent cross section, making it less efficient than in the constant case.
\citet{despali2026} finds that, for the same reason, the most massive haloes form a dark matter cored profile in the SIDM1 model, even in the presence of baryons, while vSIDM profiles remain cuspy, providing a physical pathway connecting DM physics to black hole accretion and feedback.
The HI distribution in haloes is typically very clumpy, due to the fact that HI can only survive in high-density self-shielded regions (see e.g. Fig.~\ref{fig:halo67_proj}). However, the AGN feedback can destroy such self-shielded regions and, therefore, reduce the clumpiness of gas. Moreover, the clumpiness of gas can be important in modelling the neutral gas signal. Therefore, we quantify clumpiness through the clumping factor, defined as 
$C = \langle \rho^2 \rangle / \langle \rho\rangle^2$, 
where $\rho$ is the relevant density, either total gas or HI in our case. 
In Fig.~\ref{fig:clumping}, we report the radial profiles of the clumping factor for a broader halo mass range, $M_{\rm vir}\in[10^{13.5},10^{14.5})\,\mathrm{M}_\odot$, instead of $M_{\rm vir}\sim10^{14}\,\mathrm{M}_\odot$, to obtain robust statistics (top and middle rows show the total gas and HI results, respectively).
In the bottom row of the figure, we also show the gas temperature profiles.

The gas clumping is very small for all models and only increases significantly above $C_\mathrm{gas} = 1$ close to the virial radius, implying a smooth distribution.
Interestingly, in the full samples, both SIDM1 and vSIDM haloes show slightly stronger gas clumping than the other DM models within $0.01 \lesssim \hat{r} \lesssim 0.04$ for SIDM1 and $0.02 \lesssim \hat{r} \lesssim 0.04$ for vSIDM.
In the high-$\mathrm{E}_\mathrm{tot,AGN}$ sample, the gas distribution is almost homogeneous in all four DM models due to larger AGN energy injection, whereas the gas remains significantly more clumpy in low-$\mathrm{E}_\mathrm{tot,AGN}$ sample in all models, consistent with the physical picture above. 
When turning to the HI, the overall picture holds, but now the clumping factor rises sharply with radius.
Finally, the temperature profiles of SIDM1 haloes in the full sample show colder central gas than the other models, consistent with their higher clumping factors.

\begin{figure*}
    \centering
    \includegraphics[width=\textwidth]{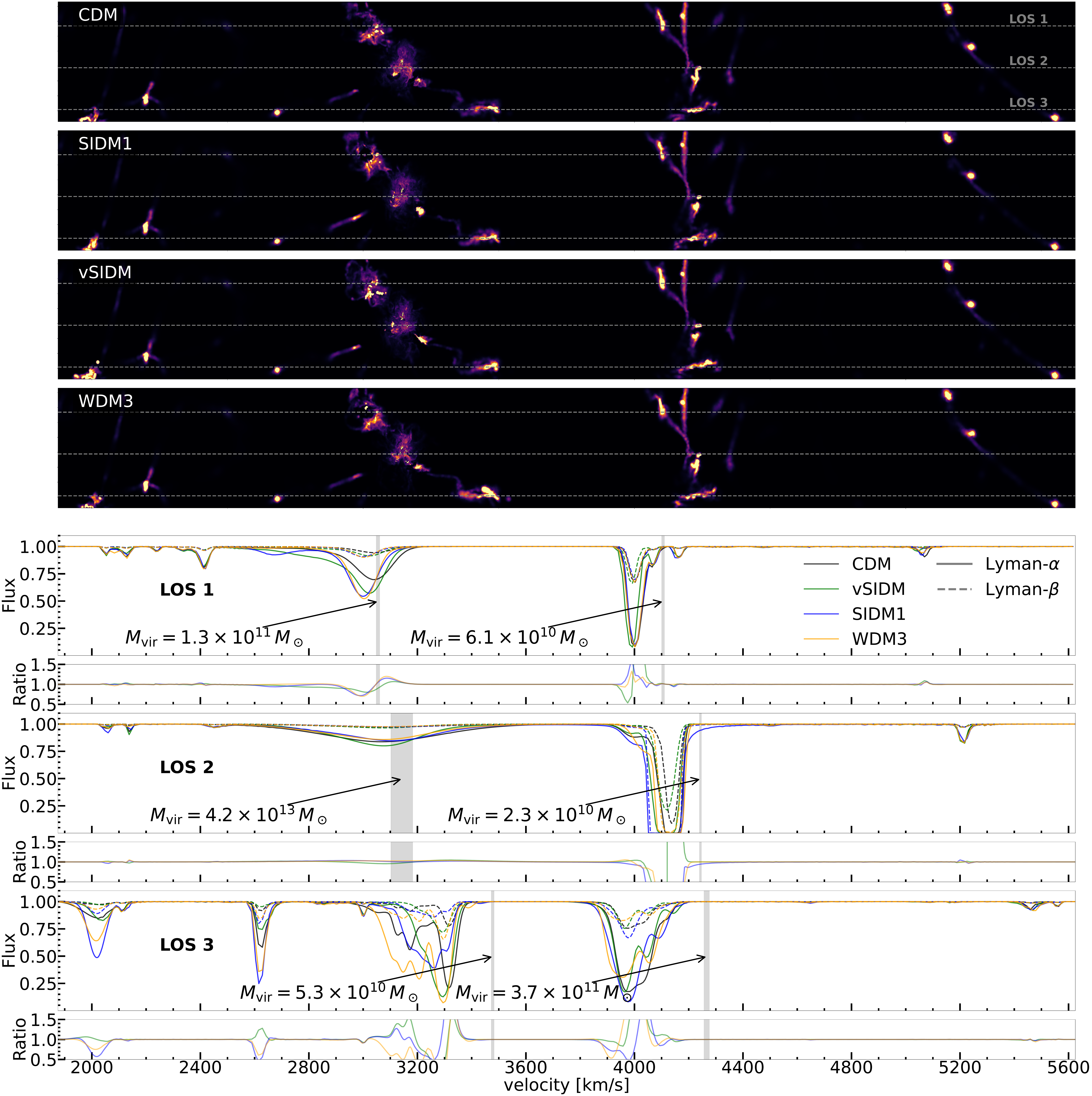}
    \caption{Upper panels: The projected HI density fields for a slice through the simulation for CDM, SIDM, vSIDM, and WDM3 (from top to bottom). 
    The slice extends the half box length of the simulated universe ($34.5\,\mathrm{cMpc}/h$), has a width of $\sim 4\,\mathrm{cMpc}/h$, and a thickness of $\sim 0.08\,\mathrm{cMpc}/h$. 
    Numbered horizontal grey dashed lines mark the lines of sight for the \lya spectra shown below.
    Lower panels: The \lya (solid) and Lyman-$\beta$ (dashed) normalised spectra along each sight line, together with their ratios to the CDM case, for each DM model (CDM: black; SIDM1: blue; vSIDM: green; WDM3: orange). 
    The spectra span the full box length (100/A) within the slice. 
    From the top to the bottom panel we show the three lines of sight indicated in the upper panels; the shaded band denotes the halo’s projection to the $x$-axis. The masses of the haloes intersected by each line of sight are also indicated.
    The DM physics alters the IGM distribution and leaves imprints on the \lymana and \lymanb spectra.
    }
    \label{fig:slice_Los}
\end{figure*}

\section{Cross-correlation between haloes and Lyman-$\alpha$ forest}
\label{sec:CCF}
A prime tool for observationally constraining DM models is the \lymana forest \citep[e.g.][]{viel05,Puchwein:2022wvk,Irsic:2023equ, garcia25}. Typically, this is done by comparing the observed \lymana forest flux along (portions of) sightlines with the predicted distribution for different ADM models. In this Section, we move beyond such global characterisation and study the cross-correlation between the \lymana forest flux and the location of haloes. 

In order to build synthetic \lymana and \lymanb spectra from the simulations, we extract line-of-sight information using the \texttt{COLT}\footnote{\url{https://colt.readthedocs.io/en/latest/index.html}} code \citep[][]{2015MNRAS.449.4336S}. \texttt{COLT} is able to reconstruct the Voronoi tessellation used by \arepo and therefore recover the simulated gas properties at the native resolution of the simulations avoiding interpolations or deposition on a grid. 
For each line of sight, we then compute the \lymana and \lymanb fluxes employing the approximation of \citet{1948ApJ...108..112H} and \citet{2006MNRAS.369.2025T} to the full Voigt-Hjerting line profile \citep{1938ApJ....88..508H}. 
We include thermal line broadening and gas peculiar velocities. The spectral resolution is $\Delta \varv = 1 \, \mathrm{km\,s}^{-1}$.

\subsection{The DM imprints on \lymana and \lymanb spectra}
We begin by building intuition on the imprints of DM physics on the \lymana and \lymanb spectra. The top panels of Fig.~\ref{fig:slice_Los} show the projected HI density fields in the same slice of the 100/A simulation runs (each panel corresponds to a different DM model). The slice extends half the box length ($37.5\,\mathrm{cMpc}/h$), with a width of $4\,\mathrm{cMpc}/h$ and a thickness of $0.08\,\mathrm{cMpc}/h$. 
We then extract three lines of sight passing through the slice (marked by horizontal grey lines in the figure) and show the corresponding \lymana (solid) and \lymanb (dashed) spectra for each DM model in the lower panels, together with their ratios relative to the CDM case.  
Along the sightlines, the locations and virial radius of the haloes responsible for the main absorption features are indicated by grey shaded bands, with their virial masses labelled.
Notably, the positions of the absorption features do not exactly coincide with the halo locations: due to peculiar velocities, the absorption is shifted in the observed spectra. Moreover, \lymanb absorption is globally weaker than \lymana because of its smaller oscillator strength. 

The absorption features produced by the same structures appear somewhat different in the different DM models, most notably near $4100\,\mathrm{km/s}$ in LOS-2 and around $3200\,\mathrm{km/s}$ in LOS-3. 
The feature at $4100\,\mathrm{km/s}$ arises from a \lymana forest absorber with HI column density $N_{\rm HI} \sim 4.9\times10^{14}\,\mathrm{cm}^{-2}$, while the broader feature at $3200\,\mathrm{km/s}$ originates from an extended tail-like structure. This appears more sensitive to the DM models, with e.g. WDM3 displaying a more extended tail, consistent with the findings of \citet{Mocz:2019pyf} and \citet{Richardson:2021onm}.

Unfortunately, the individual differences found in the \lymana forest features are strongly affected by the underlying gas configuration (i.e. it is possible to reproduce the signal found in one DM model by changing the gas distribution in another DM model), and therefore dominated by halo-to-halo scatter at fixed DM model. This motivates a statistical analysis, which we perform next.

\subsection{The galaxy-\lya cross-correlation function}
In order to statistically study the different ADM models, we use the galaxy-\lya cross-correlation function \citep[\galacc, ][]{Adelberger:2002qp,Adelberger:2005eg}, which is the cross-correlation between the position of galaxies and the \lya transmitted flux. 
This has been fruitfully employed to study quasar environments \citep[e.g. ][]{Rudie:2012mu, Prochaska:2013zea, Turner:2014dza}, metal absorption systems \citep[e.g.][]{2011A&A...530A..57S, 2017MNRAS.469L..53G, 2017ApJ...849L..18C} and reionisation \citep[e.g.][]{Kakiichi:2018yye, 2019MNRAS.483...19M, 2022MNRAS.512.4909G,2025OJAp....851666G,2023ApJ...950...66K,2025arXiv250603121K,2025arXiv250307074K,2025MNRAS.539.2790C}.

In order to compute the \galacc, we extract 30,000 lines of sight with random origins and aligned along the $x$, $y$ and $z$ axis of the simulation box (10,000 for each axis). 
Using this set of lines of sight and the halo catalogues from the simulations, we computed the \galacc $\xi(r)$ as:
\begin{equation}
\xi(r) = \frac{1}{\langle F(r) \rangle} \frac{\sum_{i \in \text {\rm pair}(r)} F_i}{N_{\text{\rm pair}}(r)},
\end{equation}
where pair$(r)$ represents all possible pairs of pixels in the synthetic spectra and galaxies in the simulation that are at distance $r$ from each other, $N_{\rm pair}(r)$ is their total number, $F_i$ is the (normalised) transmitted \lymana flux for the $i$-th pair, and $\langle F(r) \rangle$ is the average flux transmitted throughout the simulation box. 

We compute the \galacc using only the haloes in bins of virial mass defined by the following limits $[10^{8.3},\,10^9,\,10^{10},\,10^{11},\,10^{12}\,10^{13},>10^{13})\,\rm M_\odot$, at the following redshifts $z=0, 0.5, 1, 2, 3, 5$ and for all DM models investigated. For the bin $[10^{8.3},10^9)\,\rm M_\odot$ we employ the 50/A runs in order to resolve such small haloes, while for other bins we use the 100/A simulations in order to probe the largest volume available. 
We estimate the uncertainty on the \galacc by performing a bootstrapping error estimation.

\begin{figure*}
    \centering
    \includegraphics[width=\textwidth]{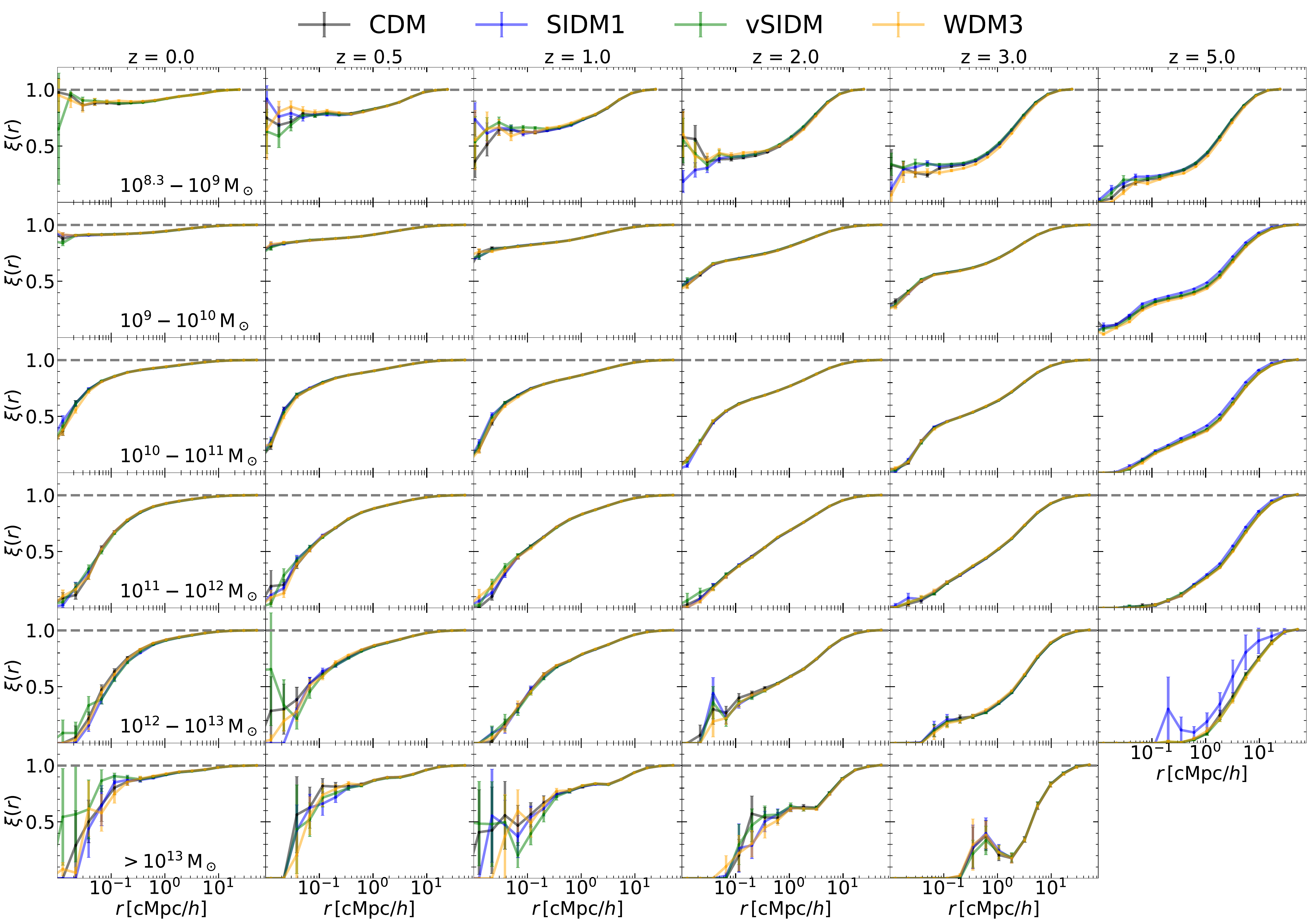}
    \caption{The \galacc for different halo-mass bins (rows) and redshifts (columns), as reported in the panels. Each curve shows results for the different DM models, namely CDM (black), SIDM1 (blue), vSIDM (green), and WDM3 (orange). The error was estimated using a bootstrap procedure. Results are computed from 100/A full-physics runs, except for the lowest-mass bin, that employed the 50/A runs in order to resolve such small haloes.
    Apart from WDM3 haloes with $10^{8.3}-10^9\,\rm M_\odot$ at $z>2$, although small differences are visible, they are within the errors set by cosmic variance and are not statistically significant.}
    \label{fig:100A_GaLaCC_all}
\end{figure*}

We show in Fig.~\ref{fig:100A_GaLaCC_all} the \galacc for all DM models as a function of increasing halo mass (from top to bottom) and redshift (from left to right). 
The resulting curves are in most cases indistinguishable for the different DM models, or with differences within the errors. There are nevertheless some potentially interesting regimes where predictions differ in the various models. For instance, the \galacc computed from the WDM3 model is systematically lower around haloes with mass $[10^{8.3},10^9)\,\rm M_\odot$ at $z>2$. However, the error associated with the \galacc is larger than the differences between different ADM models, especially at small impact parameters. This originates from the fact that small distances are not well probed because of the random nature of sightlines, which are naturally unlikely to have small impact parameters to haloes, especially larger and rarer ones. This reflects the typical approach of using quasar sightlines as background sources to measure the \lymana forest. Technological advancements and new facilities are quickly paving the way for the use of galaxies as background sources for the \lymana forest \citep[see e.g.][for initial theoretical and observational studies]{Kakiichi+2022, 2025OJAp....851666G, Meyer+2025}, thanks to their increased sensitivity and multi-object spectroscopic capabilities. For instance, facilities including WEAVE\footnote{\url{http://www.ing.iac.es/weave}}, 4MOST\footnote{\url{https://www.eso.org/sci/facilities/develop/instruments/4MOST.html}}, DESI\footnote{\url{https://www.desi.lbl.gov/}}, PFS\footnote{\url{https://subarutelescope.org/Instruments/PFS/}}, and soon ELT\footnote{\url{https://elt.eso.org/}} could enable this transformative step. In the more distant future, new-generation instruments like Wide-Field Spectroscopic Telescope (WST)\footnote{\url{https://www.wstelescope.com/}} could make this a routine task. Inspired by this possibility, we investigate a different sampling strategy.

\subsubsection{Halo-centred sightline sampling}
In order to overcome the aforementioned limitation, we now assume that the background source density is sufficient to allow us to freely choose the number and impact parameter of sightlines around a given halo. We quantify at the end of this Section whether this is feasible or not with current and future instruments. This approach allows us to better sample small halo-centric distances and maximise the information contributed by haloes, the structure affected by different DM models, to the total \galacc signal, and beat cosmic variance. Practically, for each halo, we choose sightline origins and directions in such a way to have their impact parameters follow a Gaussian distribution with standard deviation equal to a third of the median virial radius of the haloes in the same mass bin. This choice ensures that the inner halo region, where we expect the DM model to have the largest impact, is well sampled. It should be noted that, in the following, we make the assumption that the intrinsic \lymana flux can be perfectly reconstructed. While very optimistic, \citet{2025OJAp....851666G} showed that errors in such reconstruction have surprisingly small impact on the final \galacc signal. A full forward-modelling including continuum reconstruction and instrumental effects is outside the scope of this first exploratory study, given its dependence on details of the instrument used and survey configuration. We refer the reader to \citet{2025OJAp....851666G} for a thorough instrument-agnostic study of the \galacc observability under different conditions. 

\begin{figure*}
    \centering
    \includegraphics[width=\textwidth]{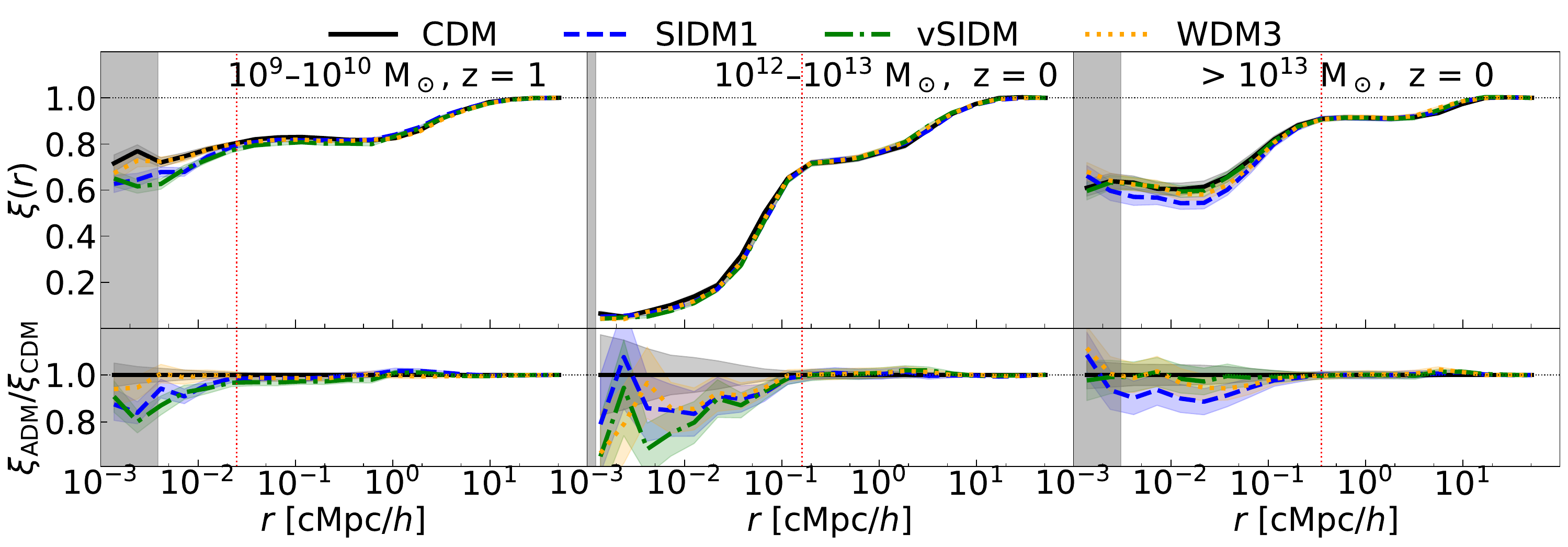}
    \caption{The \galacc computed with an optimised sightline placing (see text for details). We show the most promising combinations of halo mass and redshift. From left to right:  $10^9 \leq M_\mathrm{vir} / {\rm M_\odot} < 10^{10}$ at $z = 1$, $10^{12} \leq M_\mathrm{vir} / {\rm M_\odot} < 10^{13}$ and $M_\mathrm{vir} / {\rm M_\odot} > 10^{13}$at $z = 0$.
    The vertical red dotted lines indicate the median virial radii of the haloes, while the grey shaded regions denote the resolution limits.
    Statistically significant differences can be found within haloes, for SIDM1 and vSIDM.
    }
    \label{fig:galacc_halocentric}
\end{figure*}

We perform this alternative estimation of the \galacc for the same combinations of halo masses and redshift discussed above. For the sake of brevity, we report in Fig.~\ref{fig:galacc_halocentric} only the most interesting combinations, namely: haloes with mass $10^9 \leq M_\mathrm{vir} / {\rm M_\odot} < 10^{10}$ at $z=1$ (left panel), haloes with mass $10^{12} \leq M_\mathrm{vir} / {\rm M_\odot} < 10^{13}$ at $z=0$ (central panel) and haloes with mass $M_\mathrm{vir} / {\rm M_\odot} \geq 10^{13}$ at $z=0$ (right panel). We fix the total number of sightlines for each mass bin and DM model to $40,000$ and keep the spatial resolution of the sightline sampling the same; therefore, larger haloes are sampled by more sightlines, resulting in $\sim 1300$, $\sim 500$, and $\sim 230$ sampled haloes per mass bin (from left to right panels). 
In each panel, the horizontal and vertical dotted lines represent the average transmitted flux in the simulation and the median virial radius of the haloes used to compute the \galacc, respectively. 
The vertical grey band indicates the distance of galaxy-pixel pairs close or below the resolution limit of the simulations. 

All three benchmarks show significant differences between the DM models. 
For haloes in the bin $[10^9,10^{10})\,\rm M_\odot$ at $z=1$, SIDM1 and vSIDM exhibit lower average \lymana flux within the median virial radius, although they cannot be distinguished from each other. 
In the bin $[10^{12},10^{13})\,\rm M_\odot$ at $z=0$, vSIDM and SIDM1 show lower average \lymana flux than the other two models.
For the most massive haloes, only SIDM1 shows a different behaviour, suggesting a way to separate it from vSIDM. 
This effect arises because the effective scattering cross section in vSIDM is suppressed in more massive haloes, causing the model to approach a CDM-like regime.
We have carefully checked that these differences are not spurious by: (i) resampling the haloes and sightlines used for computing the \galacc in order to exclude outliers driving such difference, (ii) checking that the \galacc is converged both with respect to the number of haloes and to the number of sightlines per halo used, (iii) varying the number density and distribution of sightlines (namely, changing the width of the Gaussian distribution of impact parameter and by replacing it with a top hat distribution). These differences between models remain larger than the estimated uncertainties in all cases.

The galaxy formation model employed in the simulations can leave an imprint on the \galacc as large or larger than the different dark matter models, e.g., by affecting the HI distribution. We provide an estimate of this effect in Appendix~\ref{app:aida_vs_Illustris}.

We note that, if uncertainties in galaxy-formation physics can be sufficiently constrained, the \galacc has the potential to distinguish different DM models given enough sightlines. Should such a conclusion be corroborated by complementary studies, it might represent a promising way to put constraints on the nature of DM. 

Finally, we estimate the feasibility of the proposed approach by determining the minimum number of haloes and sightlines per halo required to confidently tell apart different DM models. In the most promising case, namely for haloes with mass 
$M_\mathrm{vir} / {\rm M_\odot} \geq 10^{13}$ at $z=0$, we first vary the number of randomly selected haloes, starting from $30$, while using all available sightlines for each halo, typically about $180$ per halo. This choice is motivated by the fact that the uncertainty of sightline samples is subdominant compared to the halo-to-halo variance.
We find that the models become distinguishable (at the 1$\sigma$ level) once we include $160$ haloes. We then fix the halo sample size to this value and vary the number of sightlines per halo, finding that models remain distinguishable as long as we use at least $20$ sightlines per halo. Thus we conclude that at least $160$ haloes with $\ge 20$ sightlines per halo are required to differentiate the models at 1$\sigma$. While large, these requirements are realistic for an instrument such as the WST.

\section{Conclusions}
\label{sec:conclusion}
The $\Lambda$CDM model faces some observational challenges that have prompted the investigation of alternative DM models. Understanding structure formation within these alternative DM models is a promising way to test them against observational data.
In this work, we conducted a comprehensive analysis of gas and HI properties across different dark matter models (CDM, SIDM, vSIDM, and WDM) using the AIDA-TNG simulation suite of cosmological magneto-hydrodynamical simulations. 
To connect HI properties with observational data, we examined how dark matter physics affects the gas distribution around haloes, the \lymana spectra and the correlation between \lymana spectral pixels and galaxies. 
Our main findings are as follows:
\begin{itemize}
\renewcommand\labelitemi{\bf \textendash}
    \item The gas distribution broadly follows the DM in all models, especially in high-density regions. The HI distribution is very clumpy: it correlates with gas in dense regions but less in diffuse environments, where gas is more extended than HI. 

    \item As shown in Fig.~\ref{fig:stack_100haloes},  when stacking 100 haloes with masses in the range $[10^{12.9},10^{13.1})\,\mathrm{M}_\odot$ at $z=0$, we find that DM and gas profiles form smooth halo-like profiles, while HI remains clumpy. Compared to the other DM models, SIDM1 shows a larger core in both gas and HI.
     
    \item As shown in Fig. ~\ref{fig:GAS_profiles}, the median radial gas profiles are only weakly sensitive to DM physics. 
    For haloes with $M_{\rm vir}\sim10^{13}\,\mathrm{M}_\odot$, strong feedback redistributes central gas to larger radii, producing a core that becomes more prominent towards low redshift. 
    We fit these profiles with a parametric model, with best-fit parameters listed in Appendix~\ref{tab:gas_fitting}.

    \item As shown in Fig.~\ref{fig:HI_profiles}, the median radial HI profiles become globally steeper towards low redshift, and are more feedback-sensitive than the gas.
    Differences between DM models are therefore clear, strongest for $10^{14}\,\mathrm{M}_\odot$ haloes at $z=0$, where SIDM1 preserves more HI by lowering the gas temperature and enhancing clumpiness.
    
    \item We examine AGN feedback in each DM model by splitting the $z=0$ haloes in each mass bin into the top and bottom $20\%$ in feedback strength. This split yields large HI-profile differences in all models, strongest at $M_{\rm vir}\sim10^{14}\,\mathrm{M}_\odot$, where AGN feedback efficiently ionises central HI. For the most massive haloes, feedback variations within a model produce a range of HI-depleted core sizes.

    \item For haloes in the mass range $10^{13.5}-10^{14.5}\,\mathrm{M}_\odot$, we measure the profiles for the clumpiness and for the gas temperature. In all models, clumpiness increases with radius, while a central homogeneous region grows with AGN feedback strength. SIDM is associated with lower gas temperatures, stronger clumpiness, and preserves more neutral gas.

    \item In Fig.~\ref{fig:slice_Los}, we show projected HI density in a thin slice and three \lymana/\lymanb sightlines through the slice for each DM model. 
    The spectra differ clearly between models, e.g. the strong absorption near $4100\,\mathrm{km\,s^{-1}}$ in LOS-2 and the weaker but broader feature near $3200\,\mathrm{km\,s^{-1}}$ in LOS-3, highlighting the potential of \lymana/\lymanb spectroscopy to probe DM physics.

    \item We compute the \galacc across halo masses, redshifts, and four DM models (Fig.~\ref{fig:100A_GaLaCC_all}). Differences are negligible in most cases, but WDM3 shows a potentially distinct signal for $[10^{8.3},10^{9})\,\mathrm{M}_\odot$ haloes at $z\gtrsim 3$, consistent with the suppressed halo abundance in WDM3 relative to CDM.

    \item We further probe the inner regions by sampling the environment around haloes, rather than using random lines of sight across the full box (Fig.~\ref{fig:galacc_halocentric}). SIDM1 yields a lower average \lya flux in \galacc, most notably for $[10^{9},10^{10})\,\mathrm{M}_\odot$ haloes at $z=1$ and $\gtrsim10^{13}\,\mathrm{M}_\odot$ haloes at $z=0$. For vSIDM, differences appear for $[10^{9},10^{10})\,\mathrm{M}_\odot$ at $z=1$ and $[10^{12},10^{13})\,\mathrm{M}_\odot$ at $z=0$.
\end{itemize} 

Our study shows that the impact of DM models on gas, HI, and \lya properties could be sizeable in specific mass/redshift regimes.
This opens the possibility of measuring the effect of DM through gas properties observationally. 
Furthermore, metal absorption lines are also sensitive to gas properties and could be essential for testing these predictions.  We leave these to a future work.

\section*{Acknowledgements}
CZ is supported by the China Scholarship Council for 1 year study at SISSA. EG is supported by the Kakenhi ILR 23K20035 grant. MV is supported by the INFN PD51 INDARK grant and by the INAF Theory Grant "Cosmological investigation of the cosmic web". GD acknowledges the funding by the European Union - NextGenerationEU, in the framework of the HPC project – “National Centre for HPC, Big Data and Quantum Computing” (PNRR - M4C2 - I1.4 - CN00000013 – CUP J33C22001170001). LM acknowledges the financial contribution from the PRIN-MUR 2022 (20227RNLY3) grant ``The concordance cosmological model: stress-tests with galaxy clusters'', supported by Next Generation EU, and from the grant ASI n. 2024-10-HH.0 ``Attivit\`a scientifiche per la missione Euclid - fase E''. We acknowledge ISCRA and ICSC for awarding this project access to the LEONARDO supercomputer. We acknowledge the EuroHPC Joint Undertaking for awarding the AIDA-TNG project time on the supercomputer LUMI, hosted by CSC (Finland) and the LUMI consortium, where the simulations have been run.

%%%%%%%%%%%%%%%%%%%% REFERENCES %%%%%%%%%%%%%%%%%%

%\bibliographystyle{apsrev4-2}
%\bibliography{ref} % your references Yourfile.bib
%apsrev4-2.bst 2019-01-14 (MD) hand-edited version of apsrev4-1.bst
%Control: key (0)
%Control: author (72) initials jnrlst
%Control: editor formatted (1) identically to author
%Control: production of article title (-1) disabled
%Control: page (0) single
%Control: year (1) truncated
%Control: production of eprint (0) enabled
%

%%%%%%%%%%%%%%%%%%%%%%%%%%%%%%%%%%%%%%%%%%%%%%%%%%

%%%%%%%%%%%%%%%%% APPENDICES %%%%%%%%%%%%%%%%%%%%%
\appendix
\section{A comparison with Illustris simulation}
\label{app:aida_vs_Illustris}

\begin{figure*}[ht!]
    \centering
    \includegraphics[width=\textwidth]{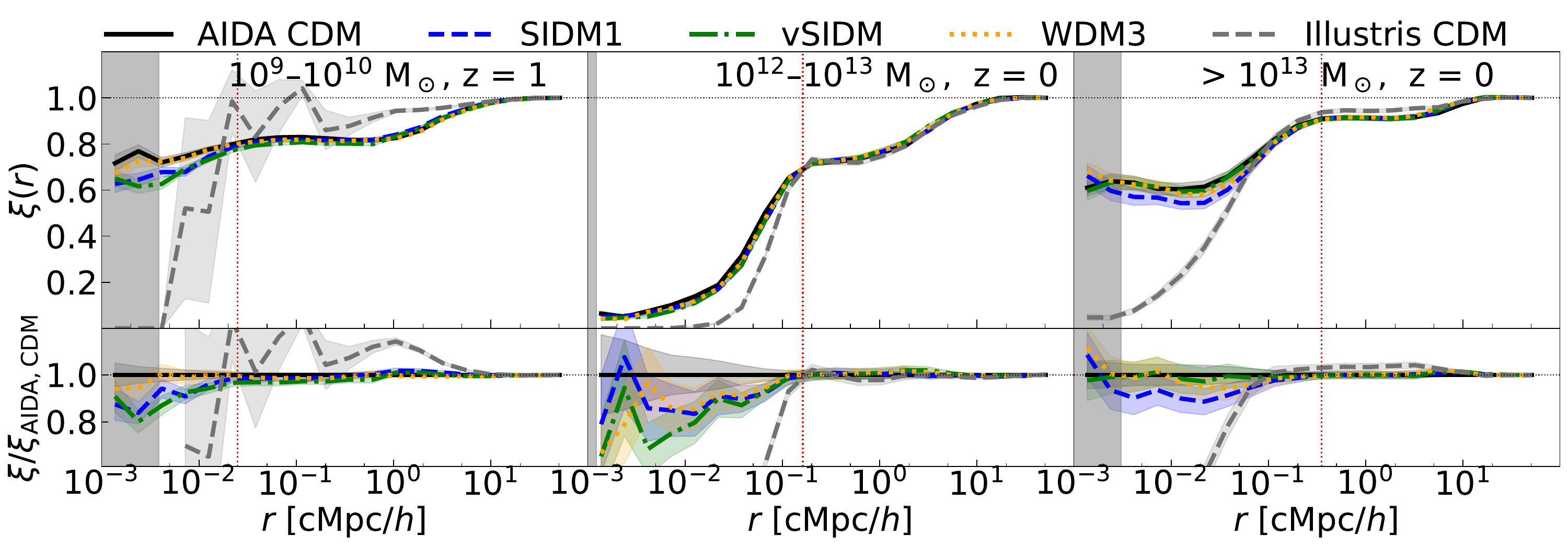}
    \caption{Same as Fig.~\ref{fig:galacc_halocentric}, with an additional (\galacc) prediction computed from Illustris-2, shown by the grey dashed curve.
    }
    \label{fig:galacc_halocentric_illustris}
\end{figure*}

To estimate the impact of different galaxy formation models, we further compare our predictions to the previous Illustris simulation\citep{vogel14,Torrey14}. In particular, we employ the Illustris-2 run, whose resolution is comparable to that of the AIDA-TNG 100/A.
As shown in Fig.~\ref{fig:galacc_halocentric_illustris}, we find that the Illustris predicts a much stronger HI absorption on the \galacc than the CDM run of AIDA-TNG, which is qualitatively consistent with previous studies of the HI distribution in haloes~\cite{2019MNRAS.486.4686K}.
In particular, \citet{2019MNRAS.486.4686K} showed that the neutral gas at large radii is preferentially aligned with the disk plane in TNG, whereas it is much more isotropically distributed in Illustris.
This difference in HI distribution provides a natural explanation for the stronger absorption we found in Illustris.
Nevetheless, such comparison should not be considered as representative of the intrinsic uncertainty of simulation-based predictions. Illustris itself shows many discrepancies with observations, that over the years have constrained its feedback model. Thus, the large difference between Illustris and AIDA-TNG does not by itself imply that the \galacc cannot distinguish DM models. Rather, it exemplifies the importance of constraining the feedback processes shaping the evolution of galaxies.

\section{Gas and HI Profiles Fits}
Here, we list the average number of haloes for the four DM models used for computing the profiles in Table~\ref{tab:prof_numhalo}.
We also present the full set of fitted parameters for the gas profiles (Table~\ref{tab:gas_fitting}) and the HI profiles (Table~\ref{tab:hi_fitting}). For each DM model, rows correspond to halo mass bins and columns to different redshift. Because of the limited number of haloes at the highest masses, we do not provide fits for the bin ($[10^{14},\infty)\,\rm M_\odot$).

\begin{table}[ht!]
\centering
\small
\setlength{\tabcolsep}{3.0pt}
\caption{The number of haloes used for computing the density profiles. Rows list halo mass for each DM model and columns are redshifts.}
\label{tab:prof_numhalo}
\begin{tabular}{c|c|cccccc}
\hline\hline
DM models & \makecell{Halo mass\\$[{\rm M}_\odot]$} & $z=0$ & $z=0.5$ & $z=1$ & $z=2$ & $z=3$ & $z=5$ \\
\hline
\multirow{6}{*}{\textbf{CDM}}
& $\sim 10^{9}$  & 786   & 1762   & 2707   & 8363  & 4665  & 4905   \\
& $\sim 10^{10}$ & 4853 & 4961 & 4985 & 4984 & 4998 & 3998 \\
& $\sim 10^{11}$ & 4909& 4979 & 4987 & 5000 & 5000 & 2352 \\
& $\sim 10^{12}$ & 971 & 993  & 1009 & 716  & 480  & 54   \\
& $\sim 10^{13}$ & 112 & 115  & 114  & 37   & 8    & ---  \\
& $\sim 10^{14}$ & 15  & 7    & 6    & ---  & ---  & ---  \\
\hline
\multirow{6}{*}{\textbf{WDM3}}
& $\sim 10^{9}$  & 420   & 1075   & 1806   & 6208  & 4566  & 4834    \\
& $\sim 10^{10}$ & 4574 & 4954 & 4982 & 4989 & 4999 & 3792 \\
& $\sim 10^{11}$ & 4915& 4978 & 4995 & 4999 & 4999 & 2348 \\
& $\sim 10^{12}$ & 970 & 993  & 1024 & 717  & 473  & 49   \\
& $\sim 10^{13}$ & 110 & 117  & 104  & 34   & 8    & ---  \\
& $\sim 10^{14}$ & 16  & 7    & 6    & ---  & ---  & ---  \\
\hline
\multirow{6}{*}{\textbf{SIDM1}}
& $\sim 10^{9}$  & --- & 1823 & 2832 & 8577 & 4705 & 4923 \\
& $\sim 10^{10}$ & --- & 4966 & 4986 & 4994 & 4998 & 4046 \\
& $\sim 10^{11}$ & 4908& 4980 & 4996 & 4998 & 5000 & 2383 \\
& $\sim 10^{12}$ & 985 & 985  & 992  & 722  & 467  & 44   \\
& $\sim 10^{13}$ & 109 & 116  & 110  & 36   & 7    & ---  \\
& $\sim 10^{14}$ & 15  & 7    & 6    & ---  & ---  & ---  \\
\hline
\multirow{6}{*}{\textbf{vSIDM}}
& $\sim 10^{9}$  & 806   & 1848   & 2892   & 8488  & 4711  & 4926   \\
& $\sim 10^{10}$ & 4898 & 4968 & 4986 & 4987 & 4999 & 4007 \\
& $\sim 10^{11}$ & 4918& 4976 & 4987 & 4999 & 4999 & 2375 \\
& $\sim 10^{12}$ & 982 & 985  & 991  & 709  & 462  & 50   \\
& $\sim 10^{13}$ & 114 & 116  & 111  & 39   & 10   & ---  \\
& $\sim 10^{14}$ & 15  & 6    & 6    & ---  & ---  & ---  \\
\hline
\end{tabular}
\end{table}

\begin{turnpage}
\begin{table*}
\centering
\caption{The fitted parameters $\Delta_0$, $r_{\rm c}$ and $\beta$ of Eq.~\ref{eq:GAS_fitting}. Rows list halo mass for each DM model and columns are redshifts. In each cell the numbers correspond to $\log_{10}(\Delta_{0}),\; \log_{10}(r_c)$\,\; $\beta$. The fits exclude profiles within the shaded (unresolved) regions shown in Fig.~\ref{fig:GAS_profiles}. Rows list halo mass for each DM model and columns are redshifts.}
\label{tab:gas_fitting}
\small
\setlength{\tabcolsep}{8pt}
\renewcommand{\arraystretch}{1.25}
\begin{tabular}{c|c|cccccc}
\hline\hline
\multicolumn{8}{c}{Parameters: $\log_{10}(\Delta_{0}),\; \log_{10}(r_c)$,\; $\beta$ ;\,Eq.~\ref{eq:GAS_fitting}}\\
\hline
\textbf{DM model} & \textbf{Halo mass [$M_\odot$]} & $z=0$ & $z=0.5$ & $z=1$ & $z=2$ & $z=3$ & $z=5$\\
\hline
\multirow{6}{*}{\textbf{CDM}}
 & $\sim 10^{9}$ & 0.35, -0.40, 0.09 & 0.70, -0.61, 0.19 & 0.70, -0.23, 0.45 & 0.99, -0.37, 0.55 & 4.90, -2.46, 0.61 & 2.94, -0.92, 0.89\\
 & $\sim 10^{10}$ & 8.00, -3.35, 0.81 & 8.09, -3.39, 0.80 & 7.93, -3.39, 0.77 & 5.30, -2.33, 0.71 & 3.64, -1.48, 0.71 & 3.29, -1.30, 0.69\\
 & $\sim 10^{11}$ & 7.59, -3.08, 0.79 & 7.36, -3.01, 0.77 & 4.69, -1.83, 0.77 & 4.35, -1.70, 0.73 & 4.08, -1.57, 0.72 & 3.98, -1.54, 0.71\\
 & $\sim 10^{12}$ & 5.21, -2.05, 0.73 & 5.51, -2.19, 0.72 & 5.54, -2.22, 0.72 & 5.57, -2.23, 0.72 & 5.44, -2.12, 0.74 & 5.54, -2.13, 0.76\\
 & $\sim 10^{13}$ & 2.78, -1.59, 0.43 & 2.98, -1.48, 0.50 & 3.22, -1.48, 0.55 & 3.28, -1.58, 0.50 & 5.10, -2.16, 0.67 & --\\
 & $\sim 10^{14}$ & 2.89, -1.67, 0.32 & 3.11, -1.38, 0.48 & 3.21, -1.60, 0.44 & -- & -- & --\\
\hline
\multirow{6}{*}{\textbf{SIDM1}}
 & $\sim 10^{9}$ & 4.68, -1.42, 0.78 & 1.68, -2.49, 0.18 & 0.71, -0.18, 0.54 & 1.05, -0.40, 0.56 & 4.99, -2.39, 0.64 & 2.95, -0.91, 0.91\\
 & $\sim 10^{10}$ & 8.68, -3.25, 0.89 & 8.31, -3.41, 0.82 & 8.05, -3.38, 0.79 & 6.32, -2.77, 0.72 & 3.76, -1.52, 0.72 & 3.40, -1.33, 0.71\\
 & $\sim 10^{11}$ & 8.22, -3.32, 0.81 & 8.13, -3.32, 0.78 & 4.75, -1.85, 0.77 & 4.53, -1.79, 0.73 & 4.21, -1.64, 0.72 & 4.04, -1.56, 0.71\\
 & $\sim 10^{12}$ & 4.90, -1.85, 0.76 & 5.29, -2.07, 0.74 & 5.29, -2.09, 0.73 & 5.24, -2.06, 0.73 & 5.25, -2.02, 0.75 & 5.49, -2.08, 0.77\\
 & $\sim 10^{13}$ & 2.67, -1.41, 0.46 & 2.93, -1.38, 0.51 & 3.12, -1.38, 0.55 & 3.10, -1.47, 0.50 & 3.56, -1.46, 0.63 & --\\
 & $\sim 10^{14}$ & 2.84, -1.42, 0.38 & 3.07, -1.34, 0.49 & 3.10, -1.43, 0.47 & -- & -- & --\\
\hline
\multirow{6}{*}{\textbf{vSIDM}}
 & $\sim 10^{9}$ & 0.29, -0.23, 0.06 & 1.69, -2.35, 0.19 & 0.71, -0.13, 0.63 & 1.03, -0.36, 0.60 & 5.01, -2.40, 0.65 & 2.96, -0.91, 0.92\\
 & $\sim 10^{10}$ & 8.28, -3.26, 0.87 & 8.39, -3.32, 0.85 & 7.93, -3.23, 0.81 & 4.05, -1.65, 0.76 & 3.51, -1.36, 0.74 & 3.21, -1.22, 0.72\\
 & $\sim 10^{11}$ & 8.37, -3.32, 0.83 & 6.14, -2.44, 0.79 & 4.53, -1.72, 0.79 & 4.28, -1.65, 0.75 & 4.06, -1.55, 0.73 & 3.96, -1.52, 0.72\\
 & $\sim 10^{12}$ & 4.99, -1.91, 0.75 & 5.34, -2.10, 0.73 & 5.28, -2.10, 0.73 & 5.27, -2.07, 0.73 & 5.32, -2.06, 0.74 & 5.54, -2.11, 0.77\\
 & $\sim 10^{13}$ & 2.70, -1.43, 0.47 & 2.96, -1.52, 0.47 & 3.06, -1.44, 0.51 & 3.28, -1.66, 0.47 & 8.20, -4.12, 0.59 & --\\
 & $\sim 10^{14}$ & 2.85, -1.43, 0.38 & 3.27, -1.66, 0.42 & 2.78, -1.02, 0.59 & -- & -- & --\\
\hline
\multirow{6}{*}{\textbf{WDM3}}
 & $\sim 10^{9}$ & 0.30, -0.24, 0.13 & 1.59, -2.45, 0.17 & 0.70, -0.37, 0.33 & 0.93, -0.33, 0.54 & 4.73, -2.39, 0.60 & 3.04, -0.95, 0.88\\
 & $\sim 10^{10}$ & 8.07, -3.32, 0.82 & 8.23, -3.42, 0.80 & 7.96, -3.40, 0.77 & 4.87, -2.11, 0.71 & 3.58, -1.42, 0.72 & 3.29, -1.29, 0.70\\
 & $\sim 10^{11}$ & 7.54, -3.04, 0.79 & 6.76, -2.74, 0.78 & 4.66, -1.80, 0.77 & 4.38, -1.70, 0.74 & 4.15, -1.59, 0.73 & 4.15, -1.60, 0.72\\
 & $\sim 10^{12}$ & 5.19, -2.05, 0.73 & 5.48, -2.17, 0.73 & 5.44, -2.15, 0.73 & 5.53, -2.20, 0.73 & 5.45, -2.11, 0.75 & 5.71, -2.16, 0.79\\
 & $\sim 10^{13}$ & 2.65, -1.45, 0.45 & 2.97, -1.59, 0.45 & 3.09, -1.51, 0.49 & 3.35, -1.60, 0.51 & 9.45, -4.33, 0.67 & --\\
 & $\sim 10^{14}$ & 2.73, -1.32, 0.39 & 3.17, -1.57, 0.42 & 2.99, -1.36, 0.47 & -- & -- & --\\
\hline
\end{tabular}
\end{table*}
\end{turnpage}

\begin{turnpage}
\begin{table*}
\centering
\caption{The fitted parameters $\Delta_0$, $r_{\rm c}$, $\beta$, $r_{\rm tr}$, $\omega$ and $\eta$ of Eq.~\ref{eq:HI_fitting}. Rows list halo mass for each DM model and columns are redshifts. Each cell shows two lines (top: $\log_{10}(\Delta_{0})$, $\log_{10}(r_c)$, $\beta$; bottom: $\log_{10}(r_{\rm tr})$, $\log_{10}(\omega)$, $\eta$). The fits exclude profiles within the shaded (unresolved) regions shown in Fig.~\ref{fig:HI_profiles}.}
\label{tab:hi_fitting}
\small
\renewcommand{\arraystretch}{1.25}
\scriptsize
\setlength{\tabcolsep}{3pt}
\resizebox{1.0\textwidth}{!}{%
\begin{tabular}{c|c|cccccc}
\hline\hline
\multicolumn{8}{c}{Per-cell ordering (two lines): $\log_{10}(\Delta_0)$, $\log_{10}(r_c)$, $\beta$ \; / \; $\log_{10}(r_{\rm tr})$, $\log_{10}(\omega)$, $\eta$ ;\,Eq.~\ref{eq:HI_fitting}}\\
\hline
\textbf{DM model} & \textbf{Halo mass [$M_\odot$]} & $z=0$ & $z=0.5$ & $z=1$ & $z=2$ & $z=3$ & $z=5$\\
\hline
\multirow{6}{*}{\textbf{CDM}}
 & $\sim 10^{9}$ & \makecell[l]{-5.48, -2.89, 0.00 \\ -2.00, -2.00, 2.00}  & \makecell[l]{-3.17, -3.00, 0.19 \\ -2.00, -2.00, 2.00}  & \makecell[l]{-2.54, -2.56, 0.28 \\ -2.00, -2.00, 2.00}  & \makecell[l]{-0.39, -2.75, 0.47 \\ -2.00, -2.00, 2.00}  & \makecell[l]{4.80, -2.40, 1.18 \\ -2.00, -2.00, 2.00}  & \makecell[l]{4.67, -0.83, 2.81 \\ -0.52, -0.95, 1.00} \\
 & $\sim 10^{10}$ & \makecell[l]{9.28, -1.09, 4.42 \\ -0.11, -1.03, 1.00} & \makecell[l]{9.30, -1.11, 4.34 \\ -0.12, -1.07, 1.00} & \makecell[l]{9.22, -1.11, 4.30 \\ -0.13, -1.06, 1.00} & \makecell[l]{5.40, -1.23, 2.26 \\ -0.07, -2.30, 1.04} & \makecell[l]{4.55, -0.98, 1.71 \\ 0.00, -4.10, 1.05} & \makecell[l]{6.47, -0.44, 5.00 \\ -0.07, -0.89, 1.00}\\
 & $\sim 10^{11}$ & \makecell[l]{4.74, -0.97, 3.19 \\ -0.07, -1.75, 1.16} & \makecell[l]{5.92, -1.09, 3.08 \\ 0.00, -0.48, 1.01} & \makecell[l]{4.38, -1.29, 2.21 \\ -1.25, -1.52, 1.79} & \makecell[l]{4.54, -0.99, 2.58 \\ -0.42, -0.81, 1.09} & \makecell[l]{3.57, -1.10, 1.23 \\ -0.02, -2.94, 1.45} & \makecell[l]{2.67, -1.38, 0.64 \\ -1.30, -3.09, 1.73}\\
 & $\sim 10^{12}$ & \makecell[l]{6.83, -1.37, 2.76 \\ -0.82, -1.45, 1.00} & \makecell[l]{4.63, -1.36, 2.06 \\ -1.00, -1.43, 1.17} & \makecell[l]{3.35, -1.35, 1.52 \\ -1.91, -3.44, 1.60} & \makecell[l]{2.73, -1.22, 1.21 \\ -1.91, -3.75, 1.19} & \makecell[l]{2.37, -1.22, 0.94 \\ -1.83, -2.50, 1.05} & \makecell[l]{2.16, -1.43, 0.53 \\ -1.45, -1.93, 1.02}\\
 & $\sim 10^{13}$ & \makecell[l]{1.71, -1.06, 1.10 \\ -1.51, -3.64, 1.03} & \makecell[l]{1.99, -1.11, 1.09 \\ -1.86, -4.03, 1.00} & \makecell[l]{3.09, -1.63, 0.94 \\ -1.56, -1.74, 1.00} & \makecell[l]{2.18, -1.34, 0.84 \\ -1.76, -3.38, 1.99} & \makecell[l]{2.00, -1.24, 0.87 \\ -1.77, -2.26, 1.06} & --\\
 & $\sim 10^{14}$ & \makecell[l]{0.62, -1.82, 0.44 \\ -1.14, -3.33, 1.00} & \makecell[l]{1.11, -1.20, 0.76 \\ -1.92, -4.72, 1.00} & \makecell[l]{2.65, -1.82, 0.82 \\ -1.91, -3.70, 1.53} & -- & -- & --\\
\hline
\multirow{6}{*}{\textbf{SIDM1}}
 & $\sim 10^{9}$ & \makecell[l]{10.17, -1.88, 1.87 \\ -0.68, -1.28, 1.01} & \makecell[l]{-2.71, -3.00, 0.24 \\ -2.00, -2.00, 2.00} & \makecell[l]{-2.14, -2.82, 0.30 \\ -2.00, -2.00, 2.00} & \makecell[l]{-0.48, -2.47, 0.51 \\ -2.00, -2.00, 2.00} & \makecell[l]{4.91, -2.29, 1.27 \\ -2.00, -2.00, 2.00} & \makecell[l]{4.89, -0.87, 2.84 \\ -0.55, -1.02, 1.00}\\
 & $\sim 10^{10}$ & \makecell[l]{2.67, -2.55, 0.93 \\ -2.00, -2.00, 2.00} & \makecell[l]{6.19, -1.83, 2.18 \\ -2.18, -2.19, 10.70} & \makecell[l]{6.15, -1.85, 2.11 \\ -2.17, -2.18, 10.90} & \makecell[l]{8.76, -0.88, 5.00 \\ -0.13, -1.07, 1.00} & \makecell[l]{3.35, -1.01, 1.71 \\ 0.00, -3.87, 2.92} & \makecell[l]{3.62, -1.44, 0.88 \\ -1.09, -1.36, 1.00}\\
 & $\sim 10^{11}$ & \makecell[l]{6.02, -1.07, 3.33 \\ -0.09, -0.86, 1.01} & \makecell[l]{6.67, -1.12, 3.29 \\ -0.06, -0.66, 1.00} & \makecell[l]{6.02, -1.62, 2.10 \\ -1.23, -1.66, 1.00} & \makecell[l]{5.45, -1.69, 1.48 \\ 0.00, 0.68, 1.00} & \makecell[l]{5.26, -1.00, 2.18 \\ -0.14, -0.61, 1.00} & \makecell[l]{2.88, -1.51, 0.64 \\ -1.30, -3.64, 1.51}\\
 & $\sim 10^{12}$ & \makecell[l]{4.36, -1.03, 2.70 \\ -0.96, -4.39, 1.77} & \makecell[l]{4.25, -1.35, 1.88 \\ -1.14, -2.63, 1.59} & \makecell[l]{3.42, -1.43, 1.42 \\ -1.94, -2.34, 1.00} & \makecell[l]{2.74, -1.25, 1.18 \\ -1.91, -2.53, 1.00} & \makecell[l]{2.43, -1.26, 0.93 \\ -1.77, -2.35, 1.04} & \makecell[l]{3.42, -2.35, 0.52 \\ -1.40, -1.96, 1.00}\\
 & $\sim 10^{13}$ & \makecell[l]{1.93, -1.18, 1.01 \\ -1.60, -2.75, 1.00} & \makecell[l]{2.14, -1.16, 1.07 \\ -1.91, -3.46, 1.00} & \makecell[l]{2.42, -1.30, 1.01 \\ -1.91, -3.35, 1.00} & \makecell[l]{1.45, -0.94, 0.91 \\ 0.00, -3.32, 99.92} & \makecell[l]{1.91, -1.24, 0.83 \\ -2.04, -2.68, 1.00} & --\\
 & $\sim 10^{14}$ & \makecell[l]{1.29, -2.99, 0.45 \\ -2.00, -3.67, 1.00} & \makecell[l]{3.19, -2.03, 0.83 \\ -1.78, -2.69, 1.35} & \makecell[l]{1.39, -1.24, 0.90 \\ -2.08, -3.64, 1.00} & -- & -- & --\\
\hline
\multirow{6}{*}{\textbf{vSIDM}}
 & $\sim 10^{9}$ & \makecell[l]{-5.12, -1.33, 0.10 \\ -2.00, -2.00, 2.00} & \makecell[l]{-2.89, -3.00, 0.22 \\ -2.00, -2.00, 2.00} & \makecell[l]{-2.31, -2.56, 0.31 \\ -2.00, -2.00, 2.00} & \makecell[l]{0.18, -2.88, 0.51 \\ -2.00, -2.00, 2.00} & \makecell[l]{4.88, -2.29, 1.27 \\ -2.00, -2.00, 2.00} & \makecell[l]{2.73, -0.51, 3.42 \\ -0.69, -3.67, 7.89}\\
 & $\sim 10^{10}$ & \makecell[l]{9.43, -1.00, 4.88 \\ -0.11, -1.08, 1.00} & \makecell[l]{9.45, -1.03, 4.78 \\ -0.12, -1.12, 1.00} & \makecell[l]{9.32, -1.02, 4.76 \\ -0.13, -1.12, 1.00} & \makecell[l]{7.97, -0.83, 5.00 \\ -0.12, -1.06, 1.00} & \makecell[l]{4.82, -0.96, 1.79 \\ -0.02, -3.49, 1.03} & \makecell[l]{4.51, -0.82, 1.15 \\ 0.00, -1.26, 1.02}\\
 & $\sim 10^{11}$ & \makecell[l]{7.38, -0.99, 3.97 \\ -0.04, -0.85, 1.00} & \makecell[l]{7.21, -1.03, 3.83 \\ -0.09, -0.81, 1.00} & \makecell[l]{6.21, -1.61, 2.18 \\ -1.20, -1.67, 1.00} & \makecell[l]{4.01, -0.97, 2.44 \\ -0.42, -0.96, 1.45} & \makecell[l]{4.60, -1.13, 1.23 \\ -0.02, -2.86, 1.04} & \makecell[l]{2.73, -1.40, 0.65 \\ -1.30, -4.18, 1.59}\\
 & $\sim 10^{12}$ & \makecell[l]{3.73, -0.95, 2.68 \\ 0.00, -4.18, 84.06} & \makecell[l]{3.50, -1.23, 1.81 \\ -2.05, -4.72, 11.07} & \makecell[l]{3.39, -1.43, 1.38 \\ -1.91, -3.44, 1.60} & \makecell[l]{2.73, -1.24, 1.18 \\ -1.92, -2.51, 1.00} & \makecell[l]{2.45, -1.28, 0.92 \\ -1.79, -2.38, 1.03} & \makecell[l]{4.48, -3.00, 0.52 \\ -1.40, -1.99, 1.00}\\
 & $\sim 10^{13}$ & \makecell[l]{2.04, -1.16, 1.09 \\ -1.60, -2.75, 1.00} & \makecell[l]{2.10, -1.21, 1.01 \\ -1.74, -3.00, 1.00} & \makecell[l]{2.12, -1.20, 1.01 \\ -1.91, -2.95, 1.00} & \makecell[l]{3.10, -1.81, 0.80 \\ -1.58, -1.97, 1.00} & \makecell[l]{1.96, -1.31, 0.79 \\ -1.84, -2.19, 1.00} & --\\
 & $\sim 10^{14}$ & \makecell[l]{-0.08, -1.50, 0.40 \\ -0.99, -3.22, 1.00} & \makecell[l]{1.32, -1.40, 0.71 \\ -2.12, -4.30, 1.00} & \makecell[l]{0.59, -0.77, 1.13 \\ -1.99, -3.66, 1.00} & -- & -- & --\\
\hline
\multirow{6}{*}{\textbf{WDM3}}
 & $\sim 10^{9}$ & \makecell[l]{-5.42, -2.92, 0.00 \\ -2.00, -2.00, 2.00} & \makecell[l]{-3.04, -3.00, 0.20 \\ -2.00, -2.00, 2.00} & \makecell[l]{-2.24, -3.00, 0.27 \\ -2.00, -2.00, 2.00} & \makecell[l]{-0.97, -2.50, 0.44 \\ -2.00, -2.00, 2.00} & \makecell[l]{4.97, -2.34, 1.23 \\ -2.00, -2.00, 2.00} & \makecell[l]{4.76, -0.95, 2.40 \\ -0.61, -1.05, 1.00}\\
 & $\sim 10^{10}$ & \makecell[l]{9.33, -1.05, 4.57 \\ -0.11, -1.05, 1.00} & \makecell[l]{9.35, -1.09, 4.42 \\ -0.12, -1.08, 1.00} & \makecell[l]{9.25, -1.09, 4.37 \\ -0.12, -1.07, 1.00} & \makecell[l]{8.32, -0.87, 4.89 \\ -0.12, -1.01, 1.00} & \makecell[l]{4.16, -0.93, 1.82 \\ 0.00, -3.47, 1.13} & \makecell[l]{6.21, -0.42, 5.00 \\ -0.06, -0.87, 1.00}\\
 & $\sim 10^{11}$ & \makecell[l]{4.56, -0.96, 3.23 \\ -0.09, -1.55, 1.26} & \makecell[l]{5.90, -1.07, 3.15 \\ 0.00, -0.50, 1.02} & \makecell[l]{5.26, -1.42, 2.21 \\ -1.15, -1.42, 1.00} & \makecell[l]{5.21, -1.54, 1.56 \\ 0.00, 0.30, 1.00} & \makecell[l]{3.42, -1.10, 1.24 \\ -0.03, -4.21, 1.67} & \makecell[l]{2.73, -1.43, 0.63 \\ -1.30, -3.14, 1.57}\\
 & $\sim 10^{12}$ & \makecell[l]{6.93, -1.40, 2.70 \\ -0.85, -1.49, 1.00} & \makecell[l]{3.42, -1.18, 2.04 \\ -1.52, -3.63, 100.00} & \makecell[l]{3.35, -1.34, 1.56 \\ -1.96, -2.45, 1.00} & \makecell[l]{2.72, -1.19, 1.26 \\ -1.91, -4.00, 1.17} & \makecell[l]{2.45, -1.27, 0.93 \\ -1.77, -2.36, 1.04} & \makecell[l]{2.23, -1.57, 0.52 \\ -1.46, -2.01, 1.03}\\
 & $\sim 10^{13}$ & \makecell[l]{1.96, -1.21, 1.04 \\ -1.60, -2.93, 1.00} & \makecell[l]{2.18, -1.27, 0.97 \\ -1.75, -3.12, 1.00} & \makecell[l]{2.27, -1.34, 0.91 \\ -1.80, -2.11, 1.00} & \makecell[l]{2.38, -1.41, 0.85 \\ -1.88, -2.38, 1.00} & \makecell[l]{1.95, -1.22, 0.83 \\ -1.76, -2.01, 1.00} & --\\
 & $\sim 10^{14}$ & \makecell[l]{-0.47, -0.66, 0.74 \\ -0.66, -2.31, 1.00} & \makecell[l]{1.38, -1.35, 0.76 \\ -2.12, -4.22, 1.00} & \makecell[l]{1.28, -1.32, 0.81 \\ -2.03, -3.50, 1.00} & -- & -- & --\\
\hline
\end{tabular}}
\end{table*}
\end{turnpage}
%%%%%%%%%%%%%%%%%%%%%%%%%%%%%%%%%%%%%%%%%%%%%%%%%%

% Don't change these lines
% \bsp	% typesetting comment
%\label{lastpage}
\end{document}